\documentclass[longauth]{aa}
\usepackage[utf8]{inputenc}

\usepackage{graphicx,caption,siunitx,xurl}

\usepackage{multirow}

\usepackage{txfonts}

\usepackage[switch]{lineno}

\usepackage[pdftex, colorlinks=true, linkcolor=blue, citecolor=blue, urlcolor=blue]{hyperref}
\newcommand\orc[1]{\href{https://orcid.org/#1}{\includegraphics[width=3mm]{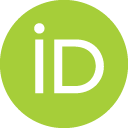}}}

\begin{document}

   \subtitle{A new class of white dwarf merger remnants with X-ray emission}

   \title{Circumstellar interaction in the extreme white dwarf merger remnant ZTF\,J1901+1458}
   \author{Aayush Desai\thanks{Corresponding author Aayush Desai: adesai@ist.ac.at}\inst{1}\orc{0009-0008-9877-5512} \and Ilaria Caiazzo\inst{1,2} \and Stephane Vennes\inst{3} \and Adela Kawka\inst{4} \and Tim Cunningham\inst{5} \and Gauri Kotiwale\inst{1} \and
   Andrei A. Cristea\inst{1} \and John C. Raymond\inst{5} \and Maria Camisassa \inst{19}\and Leandro G. Althaus\inst{18} \and J. J. Hermes\inst{7} \and Iris Traulsen\inst{19}
        \and James Fuller\inst{8}
        \and Jeremy Heyl\inst{6}
        \and Jan van Roestel\inst{9}
        \and Kevin B. Burdge\inst{10}
        \and Antonio C. Rodriguez\inst{2}
        \and Ingrid Pelisoli\inst{11}
        \and Boris T. G\"ansicke\inst{11}
        \and Paula Szkody\inst{13}
        \and Sumit\,K. Maheshwari\inst{16}
        \and Zachary P. Vanderbosch\inst{2}
        \and Andrew Drake\inst{2}
        \and Lilia Ferrario\inst{3}
        \and Dayal Wickramasinghe\inst{3}
        \and Stephen Justham\inst{14}
        \and Ruediger Pakmor\inst{14}
        \and Kareem El-Badry\inst{2}
        \and Thomas Prince\inst{2}
        \and S. R. Kulkarni\inst{2}
        \and Matthew J. Graham\inst{2}
        \and Ben Rusholme\inst{15}
        \and Russ R. Laher\inst{15}
        \and Josiah Purdum \inst{16}
        }
   \institute{
        Institute of Science and Technology Austria,
        Am Campus 1, 3400, Klosterneuburg, Austria
        \email{adesai@ist.ac.at}
        \and Division of Physics, Mathematics and Astronomy, California Institute of Technology, Pasadena, CA91125, USA
        \and Mathematical Sciences Institute, Australian National University, Hanna Neumann Building 145, ACT2601, Canberra, Australia
        \and International Center for Radio Astronomy Research, Curtin University, GPO Box U1987, Perth, WA 6845, Australia
        \and Center for Astrophysics — Harvard \& Smithsonian, 60 Garden St., Cambridge, MA 02138, USA
        \and Department of Physics and Astronomy, University of British Columbia, Vancouver, BC V6T 1Z1, Canada
        \and Department of Astronomy, Boston University, 725 Commonwealth Ave., Boston, MA 02215, USA
        \and TAPIR, Mailcode 350-17, California Institute of Technology, Pasadena, CA 91125, USA
        \and Anton Pannekoek Institute for Astronomy, University of Amsterdam, NL-1090 GE Amsterdam, the Netherlands
        \and Department of Physics, Massachusetts Institute of Technology, Cambridge, MA 02139, USA
        \and Department of Physics, University of Warwick, Gibbet Hill Road, Coventry CV4 7AL, UK
        \and University of Washington, Department of Astronomy, Box 351580, Seattle, WA 98195, USA
        \and Max-Planck-Institut für Astrophysik, Karl-Schwarzschild-Str 1, D-85748 Garching, Germany
        \and IPAC, California Institute of Technology, 1200 E. California Blvd, Pasadena, CA 91125, USA
        \and Caltech Optical Observatories, California Institute of Technology, Pasadena, CA  91125, USA
        \and Hamburger Sternwarte, University of Hamburg, Gojenbergsweg 112, 21029 Hamburg, Germany
        \and Grupo de Evolución Estelar y Pulsaciones, Facultad de Ciencias Astronómicas y Geofísicas, Universidad Nacional de La Plata, CONICET-IALP, Paseo del Bosque s/n, 1900 La Plata, Argentina
        \and Departament de Física, Universitat Politècnica de Catalunya, c/Esteve Terrades 5, 08860 Castelldefels, Spain
        \and Leibniz-Institut f\"ur Astrophysik Potsdam (AIP), An der Sternwarte 16, 14482 Potsdam, Germany
    }
   \date{}
   \titlerunning{Circumstellar interaction in ZTF\,J1901+1458}

\abstract{Double-degenerate white dwarf (WD) merger remnants can exhibit extreme magnetic fields exceeding $10^{8}$\,G and rapid rotation, but their spectral energy distributions and high-energy emission mechanisms remain poorly characterised. ZTF\,J1901+1458 stands out as the most compact and strongly magnetised object discovered in this class to date. Intriguingly, recent \textit{Chandra} observations have revealed that the white dwarf is also a source of soft X-ray emission that is too bright and hard to be of photospheric origin. We analysed new phase-resolved ultraviolet (UV) spectroscopy from the Hubble Space Telescope, together with optical and near-infrared photometry and spectroscopy, using new magnetic atmosphere models to determine its effective temperature, radius, mass, average surface magnetic-field strength, and cooling age. The spectral break at $\approx$\,3000\,$\AA$, observed in several highly magnetised WDs, is well reproduced by our new models, which account for the effect of magnetic opacities on the atmospheric structure. Our best-fit parameters for the WD yield a cooler effective temperature (T$_{\mathrm{eff}}=27,445^{+680}_{-1390}$\,K) and a larger radius than previously reported. Furthermore, the near-infrared data exclude the presence of a stellar or brown dwarf companion hotter than $\approx$\,700\,K. We jointly analysed published \textit{Chandra}/Advanced CCD Imaging Spectrometer Imaging array (ACIS-I) data and new \textit{XMM-Newton}/European Photon Imaging Camera (EPIC) X-ray spectra. The faint X-ray emission, $L_X =(1.44\pm0.13)\times10^{27}$\,erg\,s$^{-1}$, is highly pulsed at the rotation period of the WD, and the soft spectrum can be modelled by a power-law model with photon index $\Gamma=2.43^{+0.17}_{-0.15}$. We suggest that the X-rays are powered by accretion or by interaction between the WD magnetosphere and circumstellar material. A rapidly rotating magnetic field could power a weak wind along open field lines and extract material from the surface of the WD. Alternatively, low-level accretion of fallback material from the past merger event or the tidal disruption of a planetary body could supply the circumstellar material.}
\maketitle
\nolinenumbers
\section{Introduction}

White dwarfs (WDs) are the compact remnants left behind by stars less massive than about $8\,\rm{M}_\odot$. Many WDs are found in compact binaries \citep{burdge2019,burdge2020,inight2021,vanroestel2025,adamane2025,munday2025,chickles2026}, and when two WDs spiral into each other and merge, they can produce highly magnetised, rapidly rotating remnants with high magnetic fields \citep[see e.g.][]{wickramasinghe_magnetism_2000,tout2008binary,garcia2012double,schwab2021evolutionary,pakmor_large-scale_2024,munday2025-NatAstr,munday2025}. To date, however, only a handful of objects exhibiting high mass, strong magnetic fields and rapid rotation—three probable signatures of a merger origin—have been identified \citep{vennes_multiwavelength_2003,Caiazzo_2021,jewett_massive_2024,sahu2025-daq}.

The study of these merger products provides insights into magnetic-field amplification, rotational evolution, accretion processes, and binary evolution. The strong field threading the surface of these WDs affects the structure and opacities of their atmospheres and might affect their evolution; therefore, atmosphere models that account for the effects of the strong field are required to model these sources and understand their physical properties. Previous studies of extremely magnetised WDs ($B>$\,100\,MG) have shown that their spectral energy distributions (SEDs) differ from those of non-magnetised WDs, especially in the ultraviolet range \citep{green1981,schmidt1986,gaensicke2001}.

Another key factor in the evolutionary history of merger remnants is fallback accretion. Numerical simulations of double WD mergers show that right after the merger, the remnant is surrounded by a thick disc and extended tidal tails, and while some of the material is either ejected or rapidly accreted, material on eccentric orbits is accreted on longer timescales, supplying a low-level but sustained accretion flow \citep[see e.g.][]{guerrero2004,rosswog2007,loren-aguilar_high-resolution_2009,dan2014structure}. This fallback rate is initially high, but decays over time, releasing gravitational potential energy that can heat the remnant’s outer layers and power high-energy emission. Such a mechanism may also affect the spin evolution and thermal structure of the remnant over extended timescales and would provide another mechanism to identify merger remnants \citep[see e.g.][]{rueda_electromagnetic_2019,sousa_double_2022,yang_revealing_2022}. However, until recently, no signs of long-term accretion from fallback have been observed in a merger remnant.

ZTF\,J1901+1458 (hereafter J1901) has emerged as a cornerstone object in the field of WD merger remnants. Discovered in 2021 \citep[][hereafter C21]{Caiazzo_2021} as a variable source in the Zwicky Transient Facility archive \citep{bellm2019,graham2019,dekany2020,masci_zwicky_2019}, J1901 was immediately identified as unusual. It was reported to have a small radius of only $\approx$\,2140\,km and a mass in the range 1.327–1.365\,M$_{\odot}$, with a short rotation period of 6.94 minutes. Its surface magnetic field was inferred to reach 600–900\,MG, placing it among the most strongly magnetised WDs known.

As typical WDs exhibit rotation periods of hours to days \citep[see e.g.][]{koester_search_1998,charpinet_seismic_2009,hermes_white_2017}, the rapid rotation rate of J1901, combined with its high mass and intense magnetic field, provides strong evidence that the star was born in a double-degenerate merger. Additionally, the radius and mass derived in the discovery paper imply that J1901 is close to the Chandrasekhar limit. C21 also noted that electron captures could, in principle, bring J1901 close to instability: at the high densities reached in the WD core, $^{23}$Na atoms can undergo inverse beta decay, and therefore capture an electron to produce $^{23}$Ne atoms, reducing the electron fraction $Y_e$ and degeneracy pressure.
This change could then increase the core density above the threshold for capture on $^{24}$Mg and $^{20}$Ne, further reducing $Y_e$ and making J1901 unstable to collapse. J1901 is thus a valuable case to investigate the interplay between chemical stratification, neutrino cooling, and the stability of ultra-massive WDs.

The initial characterisation of J1901, however, was based only on optical and near-ultraviolet (near-UV) data, and the estimates for the temperature and radius of the star (and thus of the mass and age) were obtained by modelling the observed SED of the WD with non-magnetic atmosphere models. As we mention above, and as we show in detail in this work, magnetic fields have an important effect on the emission from ultra-magnetised WDs, both in the location and strength of the absorption lines and in the shape of the overall SED; therefore, these estimates of J1901 should be revisited. For this reason, we undertook a comprehensive follow-up campaign of J1901. We leveraged ultraviolet observations (from the Hubble Space Telescope; details in Section \ref{sec:Observations}) alongside new optical and infrared photometry and spectroscopy to refine J1901’s fundamental parameters and constrain the presence of a stellar or substellar companion (Section \ref{sec:Results}). The inclusion of ultraviolet spectroscopy is important: at the effective temperature of J1901 ($\approx$\,27,400\,K), a significant portion of the WD’s flux emerges in the UV, thus providing a tight handle on the effective temperature, radius, and interstellar extinction that optical data alone cannot constrain well. Most importantly, for the characterisation, we employed a new set of magnetic atmosphere models that included, for the first time, the effect of magnetic opacities on the structure of the atmosphere and are able to reproduce the full SED of ultra-magnetised WDs.

Another focus of this work is the high-energy emission of J1901. The WD has been the target of two X-ray campaigns with \textit{Chandra} (proposal ID 24200244, PI Safdi) and \textit{XMM-Newton} (ObsID 0922750101, PI Traulsen) to search for emission from axion-like particles. Axions could be produced in the dense hot cores of WDs through electron bremsstrahlung \citep{ning2025} and nuclear transitions \citep{fleury2023}, and subsequently converted into high-energy photons in the WD magnetosphere \citep{dessert2019,dessert2022}. Although no evidence for axion bremsstrahlung emission was found in the \textit{Chandra} data \citep{ning2025}, the X-ray flux shows a strong excess over the expected photospheric emission based on the measured temperature and radius of the WD. We re-analysed the \textit{Chandra} data for J1901 and present the new \textit{XMM-Newton} data, after performing both timing and spectroscopic analyses (Section~\ref{sec:xrayspec}). In the absence of a mass-transferring companion, we present several alternative explanations for the origin of the X-ray emission.

Another object with strong similarities to J1901 is ZTF\,J2008+4449 (hereafter J2008), whose discovery we present in a companion paper \citep{cristea2026}. J2008 is isolated, massive (1.1\,M$_\odot$), rapidly rotating (P$_{\rm rot}\!\approx\!7$ minutes), and strongly magnetised ($B_{\rm p}\approx5\times10^8$\,G). It is likewise a soft X-ray emitter, with a spectral shape similar to that of J1901, suggesting that both objects may be members of a new class of merger remnants interacting with circumstellar material.

The paper is structured as follows. Section~\ref{sec:Observations} describes the observations, Section~\ref{sec:Results} presents the timing and spectroscopic analysis, and Section~\ref{sec:Modelling} introduces the magnetic atmosphere models used to characterise J1901. Section~\ref{sec:Discussion} discusses the implications of our findings, including the comparison with J2008 and other merger remnants (Section~\ref{sec:Comparison}). We summarise our results and outline future prospects in Section~\ref{sec:Conclusions}.

\section{Observations \label{sec:Observations}}

We begin by presenting all available data for J1901, including newly obtained X-ray and ultraviolet (UV) spectroscopy. To properly phase the observations between different epochs, we converted all times into modified Julian date (MJD) in the barycentric dynamical time (TDB) scale.
\subsection{\textit{HST} data}
J1901 was observed \citep[programme ID 16753][]{caiazzo_HSTprop2021} with the Cosmic Origins Spectrograph (COS) and Space Telescope Imaging Spectrograph \citep{green2011cosmic,woodgate1998space} on the Hubble Space Telescope (\textit{HST}). The details of the observations are summarised in Table~\ref{tab:HST}. The low-resolution G140L grating, with a central wavelength of 800\,\AA, was used on COS, yielding spectral coverage over $\simeq$\,900--1900\,\AA\ at a resolving power of R\,$\approx$\,2000. For STIS, we used the G230L near-ultraviolet MAMA grating, covering $\simeq$\,1600--3000\,\AA\ with a resolving power of R\,$\approx$\,500. We reduced the data using the default \texttt{CalCOS} and \texttt{CalSTIS} pipelines for COS and STIS, respectively \citep{kaiser2008cos,hulbert1997stis}. Since the observations were taken in time-tag mode, we phase-binned the spectra at the 6.9-minute period using the default \texttt{splittag} and \texttt{inttag} Python functions. All timestamps were converted to the barycentric dynamical time (TDB) system using an adapted version of the \texttt{barycorrpy} routine. The effective pixels of the COS XDL (the cross-delay-line detector of the far-ultraviolet channel) highly oversample the spectral resolution, with about six bins per resolution element.\textsuperscript{\ref{fn:first},\ref{fn:second}}\stepcounter{footnote}\footnotetext[\value{footnote}]{\url{https://hst-docs.stsci.edu/cosihb/chapter-3-description-and-performance-of-the-cos-optics/3-2-size-of-a-resolution-element}\label{fn:first}}\stepcounter{footnote}\footnotetext[\value{footnote}]{The far-UV detector does not have physical pixels; digitised positions are derived for each count, and these are what we consider to be pixels.\label{fn:second}}. We rebinned the flux and its error for each phase-resolved spectrum using the \textit{SpectRes} code \citep[see][]{carnall2017spectres}.

\subsection{X-ray data}
\subsubsection{Chandra X-ray Observatory}

J1901 was observed with the ACIS-I instrument \citep{garmire2003-acis} on board the \textit{Chandra X-ray Observatory} \citep{weisskopf2000} on 9 and 10 December 2022. The campaign \citep[proposal ID 24200244;][]{safdi2022} comprised three observations (ObsIDs 26496, 27596 and 27597), with exposure times of 10--15\,ks and a combined exposure time of 39.3\,ks. All observations were performed in very faint mode, which stores event grades in regions of 5$\times$5 pixels and enables improved rejection of background events. We reduced the data using the \textit{Chandra Interactive Analysis of Observations} software package \citep[\texttt{CIAO};][]{fruscione2006-CIAO}.

\subsubsection{\textit{XMM-Newton}}

J1901 was also observed with \textit{XMM-Newton} \citep{jansen2001} on 21 and 22 March 2024 (ObsID 0922750101; PI Traulsen) for 77.8\,ks. In this work, we rely primarily on data from the European Photon Imaging Cameras (EPIC): pn \citep{struder2001-epic-pn} and MOS1/2 \citep{turner2001-epic-mos}, as well as the Optical Monitor \citep[OM;][]{mason2001-om}. Given the relatively low S/N of the EPIC data, we do not analyse the Reflection Grating Spectrometer (RGS) event data. The observation data files and Pipeline Processing System products (PPS) were retrieved from the \textit{XMM-Newton} Science Archive.\footnote{\url{https://www.cosmos.esa.int/web/xmm-newton/xsa}}

The \textit{XMM-Newton} observations also included exposures with the optical monitor in Fast Mode, using the UVW1 and UVM2 filters. The eight exposures with UVW1 had exposure times ranging from 2.5--4.4\,ks, with a collective total of 32.9\,ks. The eight exposures with UVM2 had exposure times of 4.4\,ks with a collective total of 35.3\,ks. For the analysis of the optical monitor data, we make use of the time-series data provided in the pipeline data products PPS.

\subsection{Optical photometry}
To model the SED of the WD, we employ available photometric data from the Pan-STARRS PS1 survey \citep{chambers_pan-starrs1_2019}. Since the reported errors are smaller than the intrinsic photometric variation, we use an uncertainty of 0.02 magnitudes to account for variability. These are the errors reported in Table~\ref{tab:photometry}.

We supplement our optical photometry with light curves from the ZTF survey \citep{masci_zwicky_2019,bellm2019,dekany2020,graham2019} in the $g$, $r$ and $i$ bands. The ZTF light curves are available from the ZTF archive\footnote{\url{https://irsa.ipac.caltech.edu/Missions/ztf.html}} and were extracted using the \texttt{ztfquery} Python package \citep{rigault_ztfquery_2018}.
We also acquired high-speed photometric observations in the \textit{g} and \textit{r} filters using the Caltech HIgh-speed Multi-colour camERA \cite[CHIMERA;][]{harding_chimera_2016} mounted on the 200-inch Hale Telescope at Palomar Observatory on three different nights: 18 August 2020, 24 April 2025 and 30 May 2025. The exposure time for each image was 3\,s, and the total exposure times were 80, 30 and 75 minutes, respectively. The absolute local time of each exposure was recorded with millisecond precision using a GPS receiver. Standard bias subtraction and flat-fielding procedures were applied to the images. Aperture photometry was then performed using an adapted version of the ULTRACAM pipeline \citep{dhillon2007}, with differential light curves constructed relative to a single non-variable comparison star. Finally, we obtained high-speed photometry in the \textit{u} band using \texttt{LightSpeed}, which was temporarily mounted on the Hale Telescope for testing on the night of 27 October 2024. The total exposure time was 40 minutes, while the exposure time for each image was 1\,s. The light curve was obtained by reducing the images using a bespoke reduction pipeline developed for LightSpeed.

\subsection{Infrared data}
We obtained near-infrared spectroscopy with the Folded-port InfraRed Echellette spectrograph \citep[FIRE,][]{simcoe2008,simcoe2010} on the Magellan Baade 6.5 m telescope. We observed J1901 on 15$^{\rm th}$ July 2025 in the echelle mode, which provides a resolving power of $R$\,$\approx$\,6000 spanning the 1.0–2.5 $\mu$m range. The observations were conducted with the 0.6\H\, slit, at parallactic angle to minimise differential atmospheric refraction. We used exposure times of 600\,s, for sequences of ABAB dithers across 9 individual science exposures, with a total integration of 5400\,s. We chose the up-the-ramp sampling mode for the readout in order to minimise overheads. We also obtained a standard star for telluric correction. The data were reduced using the \texttt{pypeit} package for semi-automated reduction of astronomical slit-based spectroscopy \citep{prochaska2020,prochaska2020zndo}.
We also use archival UKIDSS $J$, $H$, and $K_s$ photometry to anchor the near-infrared SED. The photometry is listed in Table~\ref{tab:photometry}, and the aperture-photometry procedure is described in Appendix~\ref{sec:appendix-data}.

\section{Analysis \label{sec:Results}}

\subsection{Analysis of the UV spectra}
\begin{figure*}[tb]
    \centering
    \includegraphics[width=1\linewidth]{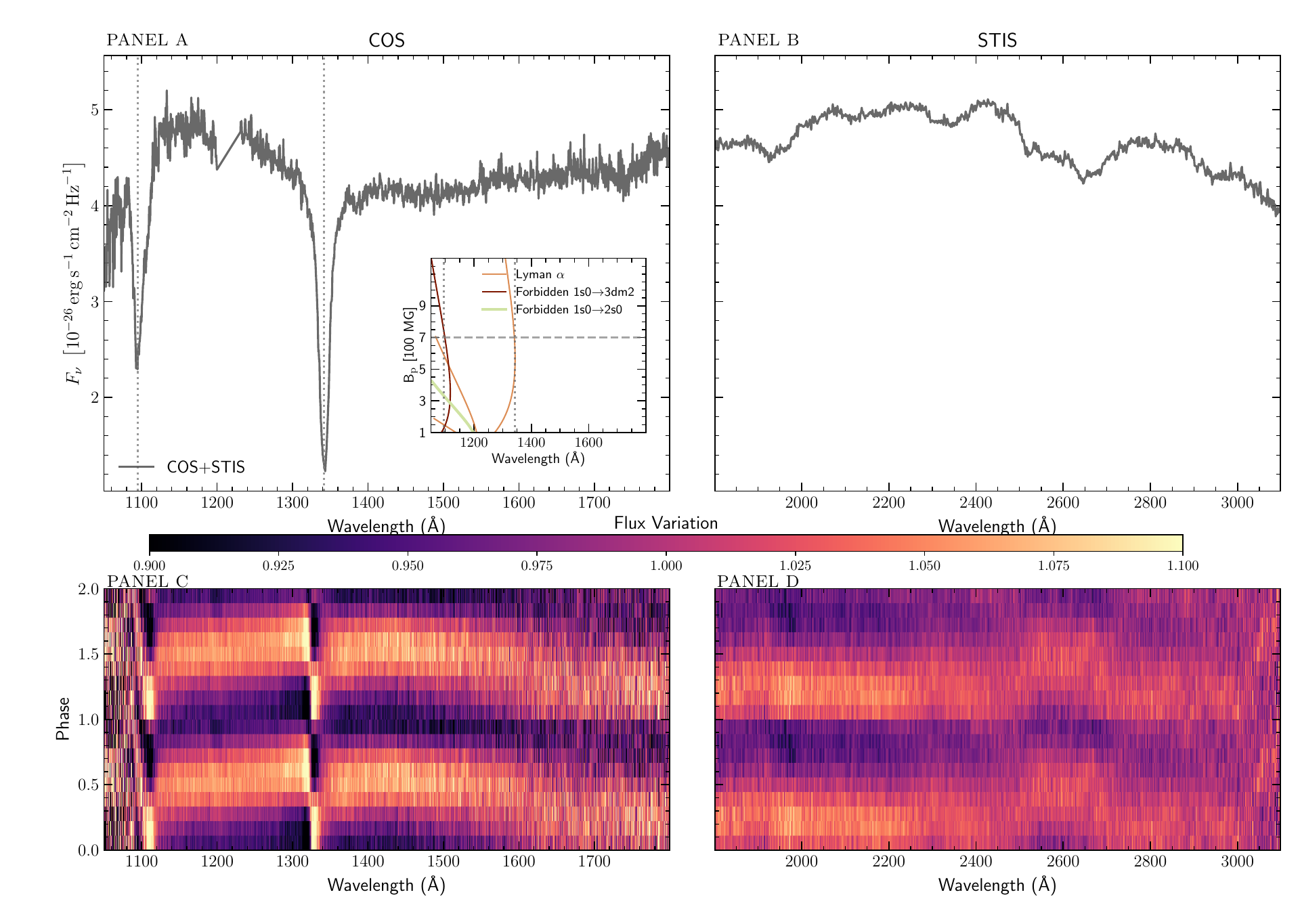}
       \caption{COS and STIS spectra of J1901. \textsc{Panel a}: Phase-averaged COS spectrum. The inset shows the expected positions of Lyman $\alpha$ and forbidden hydrogen transitions as a function of magnetic field strength; the vertical dotted lines mark the observed absorption features near 1096 and 1342\,\AA. \textsc{Panel b}: Phase-averaged STIS spectrum. \textsc{Panels c,d}: Phase-resolved COS and STIS spectra divided by the phase-averaged spectrum and plotted over two rotation cycles. The colour bar shows the fractional flux variation with respect to the mean. The spectra show a strong variation in the continuum with shifts in phase that depend on wavelength. Furthermore, the two absorption features in the COS spectrum show a similar shift in wavelength with phase (see also Fig.\,\ref{fig:ly-alpha-fit}).}
   \label{fig:COS-STIS}
\end{figure*}

The full UV spectrum is shown in Fig.\,\ref{fig:COS-STIS}, with COS in \textsc{Panel a} (1100--1800\,\AA) and STIS in \textsc{Panel b} (1800--3100\,\AA). The COS spectrum contains two strong absorption lines at 1096 and 1342\,\AA, while the STIS spectrum shows broader and weaker absorption features. The inset plot of \textsc{Panel a} compares the location of the observed absorption features to the predicted wavelengths of hydrogen absorption lines, split in Zeeman components and shifted by the strong magnetic field, as a function of magnetic field strength on the surface of the WD \citep{2014ApJS..212...26S}.
The line observed at 1342\,\AA\ corresponds to a Lyman\,$\alpha$ component (1s0\,$\rightarrow$\,2p0), while the line at 1096\,\AA\ is a blend of a permitted Ly\,$\alpha$ component (1s0\,$\rightarrow$\,2p$-1$) and a forbidden Ly\,$\beta$ component (1s0\,$\rightarrow$\,3d$-$2). Their locations indicate an average surface field strength of B$_{\rm avg}\!\simeq\!7\times10^{8}$\,G.

We binned the data into ten phases to analyse its variation over the rotation period. The phase-resolved spectra, divided by the phase-averaged spectrum and repeated over two cycles, are shown in \textsc{Panels c} and \textsc{d} of Fig.\,\ref{fig:COS-STIS}. The continuum shows a sinusoidal modulation whose phase depends on wavelength: the minimum in brightness occurs at different phases in different bands, and the peak-to-peak amplitude increases towards bluer wavelengths, from 4\% in the reddest STIS band to 12\% in the bluest COS band (see also Section~\ref{sec:UVOrigin}). We also detect wavelength shifts in the COS absorption features, as well as changes in their equivalent widths.

\begin{figure}
    \centering
    \includegraphics[width=1\linewidth]{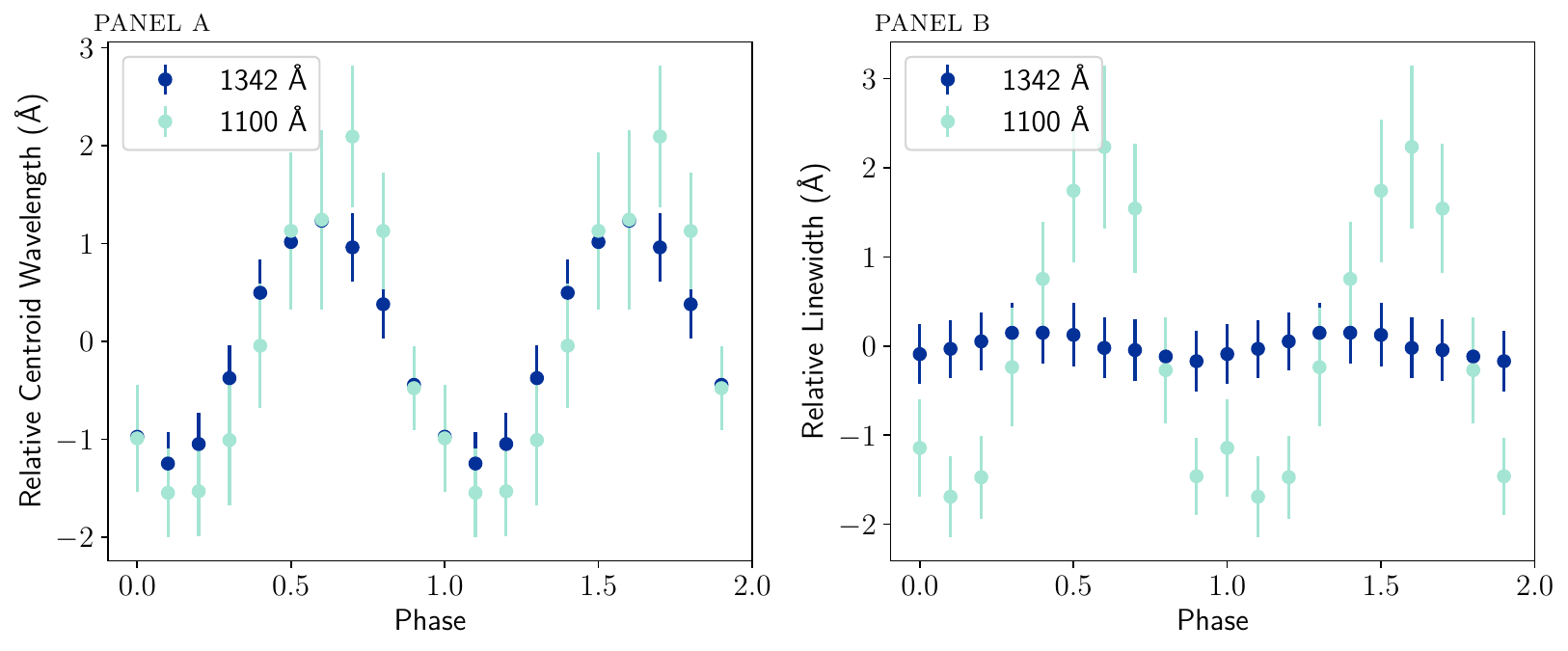}
    \caption{Phase-resolved Gaussian fits to the absorption features at $\approx$\,1096.30\,\AA\ (light blue points) and $\approx$\,1342.16\,\AA\ (dark blue points). \textsc{Panel a} shows the centroid offset from the phase-averaged value. \textsc{Panel b} shows the corresponding offset in Gaussian width. In both panels, the relative values are with respect to the mean value over the ten phase bins.}
    \label{fig:ly-alpha-fit}
\end{figure}

To quantify these line-profile variations, we perform Gaussian fits to the two strong features at 1096 and 1342\,\AA, and the results are shown in Figure~\ref{fig:ly-alpha-fit}. The centroids of the two features shift in phase with each other by $\approx$\,2\,\AA, while the width of the 1100\,\AA\ line varies more strongly and slightly out of phase with that of the 1342\,\AA\ line.

From the inset in Fig.\,\ref{fig:COS-STIS}, we can see that the 1s0\,$\rightarrow$\,2p0 Lyman $\alpha$ component and the 1s0\,$\rightarrow$\,3d$-$2 forbidden transition of Lyman $\beta$ vary with field in a similar fashion, while the $\pi$ component of the Lyman $\alpha$ transition varies much faster. It is therefore likely that the common shift in wavelength is due to a shift in magnetic field and that the 1100\,\AA\ line is dominated by the 1s0\,$\rightarrow$\,3d$-$2 forbidden transition; the stronger variation in amplitude of this feature, however, is consistent with a significant contribution from the $\pi$ component of Ly$\alpha$, which is more sensitive to the magnetic field strength in this region.

\subsection{Analysis of the X-ray data}
\label{sec:xrayspec}
\subsubsection{Reduction and detection significance}
\citet{bamba_x-ray_2024} and \citet{ning2025} previously analysed the \textit{Chandra} X-ray observations and reported marginal X-ray detections (at the $3.5\sigma$ and $2.6\sigma$ levels, respectively), finding an X-ray flux of $\approx10^{-15}$\,erg\,s$^{-1}$\,cm$^{-2}$ in the 1.0--3.0\,keV band. Both studies noted that the \texttt{wavdetect} source-detection algorithm distributed with CIAO yielded no source detection at the target. We therefore re-analysed the \textit{Chandra} data. We began by reprocessing the three observations using the \texttt{chandra\_repro} tool and used \texttt{merge\_obs} to merge them into a single event file.

The PSF and exposure maps were generated with the exposure evaluated at 0.9\,keV, approximately the median source photon energy. We extracted spectra for the source and an adjacent background region using the standard procedure outlined in the \texttt{CIAO} documentation. The source and background circular apertures had radii of 2 and 80 arcsec, respectively. Additionally, we performed source detection using \texttt{wavdetect} on images in the soft, medium, and hard energy bands. Our new analysis yielded an X-ray source at the expected target location with a significance above $6\sigma$. We examined the detection significance in the ACIS science energy bands: ultra-soft (0.2--0.5\,keV), soft (0.5--1.2\,keV), medium (1.2--2.0\,keV), hard (2.0--7.0\,keV), and broad (0.5--7.0\,keV). We found significant detections in the medium and broad bands, with significances of 6.5 and 6.9$\sigma$, respectively. The detection significance is dominated by counts in the medium band, in which we identified six source counts above an expected background of 0.06 counts in the 2 arcsec source aperture over the total 39.3\,ks exposure.

All other \textit{Chandra} science bands yielded a detection significance below $3\sigma$. These results, together with confidence limits on the source count rates determined using the Bayesian prescription of \citet{kraft-burrows-nousek-1991}, are summarised in Appendix Table~\ref{tab:chandra_acis_counts}. In the medium band, \textit{wavdetect} found a source at the target location, while no source was detected in the other two bands.

\begin{figure*}[tb]
    \centering
    \includegraphics[width=0.3\linewidth]{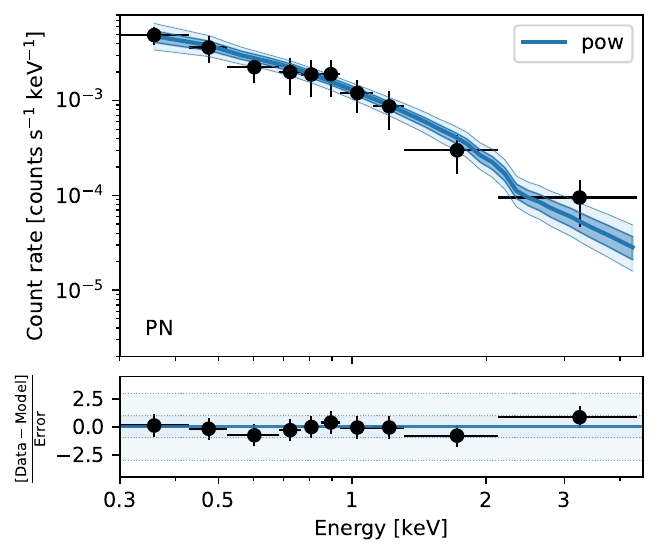}
    \includegraphics[width=0.3\linewidth]{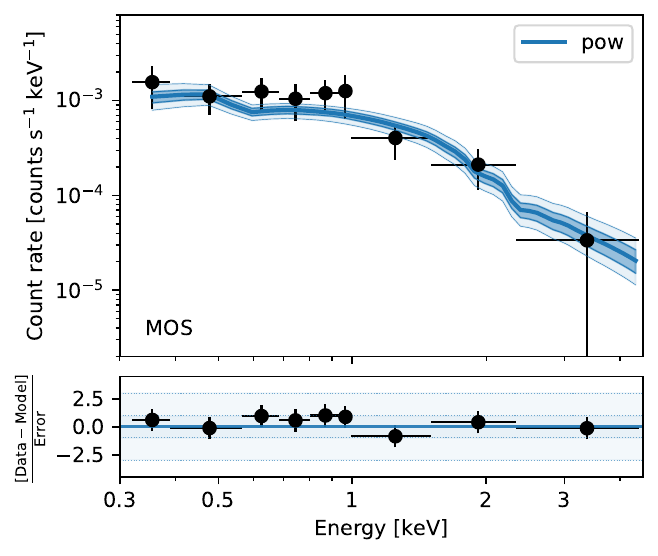}
    \includegraphics[width=0.3\linewidth]{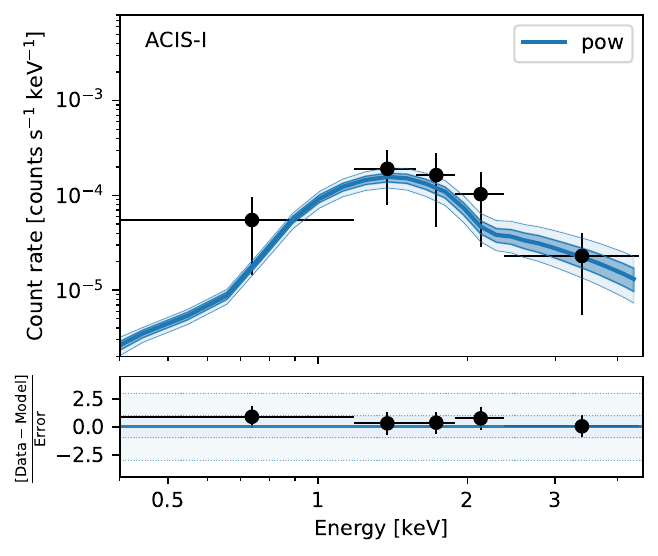}
    \includegraphics[width=0.3\linewidth]{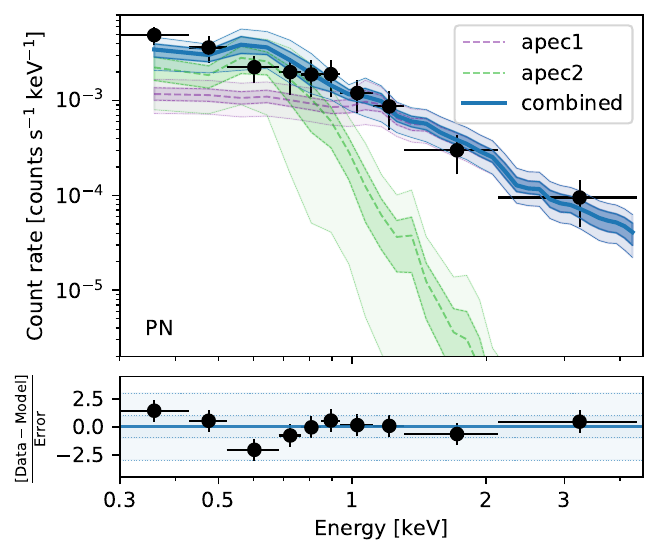}
    \includegraphics[width=0.3\linewidth]{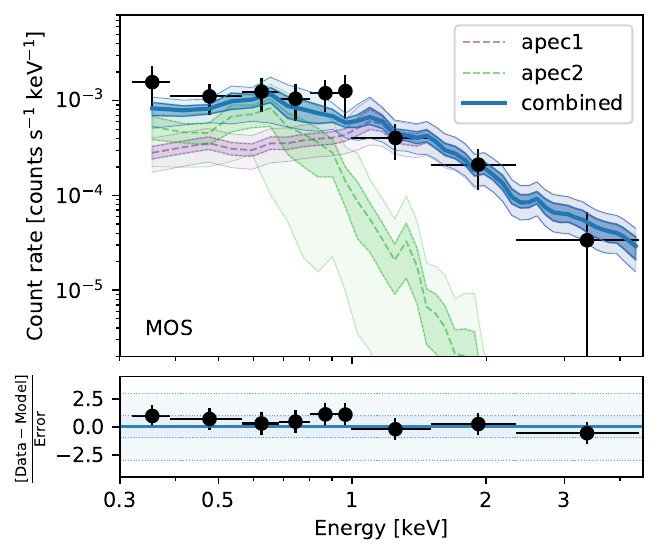}
    \includegraphics[width=0.3\linewidth]{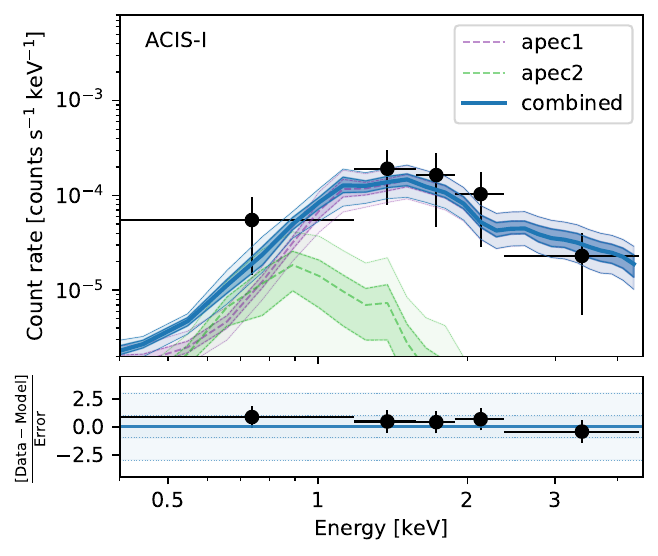}

    \caption{Fit to the available archival X-ray spectroscopic data, including \textit{XMM-Newton} EPIC pn (left), MOS (middle), and \textit{Chandra} ACIS-I (right). The top panels show a fit using an absorbed power-law model, while the bottom panels employ a two-temperature isothermal plasma model (APEC) with an absorption component. The data points show the binned X-ray spectra, while the best-fitting model is shown in blue. Contours indicate 1 and 3$\sigma$ limits on the best-fitting model, including the two individual APEC components in the bottom panels.}
    \label{fig:xspec}
\end{figure*}

For the new \textit{XMM-Newton} observations, the EPIC data were reduced using v21.0.0 of the Science Analysis Software \citep[\texttt{SAS},][]{gabriel2004-sas}. We performed a filtering of event files using the \texttt{SAS} routine \texttt{evselect} and defined Good Time Intervals (GTI) to remove contamination from high flaring background times, with thresholds on the total CCD count rate defined via visual inspection of the light curve to be 0.5, 0.15 and 0.1 counts\,s$^{-1}$ for pn, MOS1 and MOS2, respectively. This procedure resulted in total effective exposure times of 53.1, 65.1, and 67.3\,ks for pn, MOS1 and MOS2, respectively. We extracted spectra for the source and adjacent background region in each camera, with apertures of radius 20 and 80\, arcsec, respectively.  We find that the source is confidently detected in the 0.5--2.0\,keV band in all three EPIC cameras, with pn, M1, and M2 yielding detection significances of 9.3$\sigma$, 5.7$\sigma$, and 7.3$\sigma$, respectively. These detections are consistent with the expectation from the re-analysis of the \textit{Chandra} data. Aided by the significantly larger effective area at low energies ($<1.0$\,keV), the source is also detected in the 0.2--0.5\,keV band in pn (10.4$\sigma$) and M2 (8.7$\sigma$), although M1 alone does not yield a significant detection ($2.3\,\sigma$). This further confirms the soft nature of the measured X-ray spectrum. The source counts, associated 90\% confidence limits on source count rates, and detection significances are presented in Appendix Table~\ref{tab:allinst_counts}. We also analyse the time-series EPIC data to search for periodicity in the X-ray emission.\footnote{We note that the time scale in the header after barycentric correction is rightly indicated as TDB (barycentric dynamical time); however, the comment to the entry remains unchanged, i.e. XMM time will be terrestrial time (TT), and might generate confusion.}

\subsubsection{Spectral analysis}
We performed a joint \textit{Chandra} and \textit{XMM-Newton} spectral analysis using the Bayesian X-ray Analysis software package \cite[BXA;][]{buchner2016-bxa}, which connects \texttt{ULTRANEST} \citep{buchner2016-bxa}, a nested-sampling algorithm, with an X-ray spectral-fitting environment. We used \texttt{XSPEC} \citep{arnaud1996XSPEC} through its \texttt{PYTHON} interface, \texttt{PYXSPEC}. We simultaneously fitted the \textit{Chandra} ACIS-I and \textit{XMM-Newton} EPIC pn and MOS data using the Cash statistic, which is necessary for accurately fitting data in the low-count regime \citep{cash1979}. Before fitting, the spectra were binned using the optimal-binning algorithm of \citet{kaastra2016-optimal-binning}. As the X-ray emission mechanism is unknown, we used different models to explore both a non-thermal scenario, with a power law, and thermal emission, with either a blackbody or an optically thin plasma model (\texttt{APEC}), which is often used to model the X-ray emission of accreting magnetic WDs. For the \texttt{APEC} model, we adopt solar abundances as presented by \citet{asplund2009}. At a distance of 41\,pc, the system is sufficiently close to Earth that the Galactic $n_{\rm H}$ absorption should be minimal, so we do not include any in the models.

We impose a uniform prior on the photon index for the power-law model and on the temperature for the blackbody and the optically thin plasma model, while we adopt log-uniform priors for the normalisation. Finally, we make use of the model comparison framework within \texttt{BXA} \citep[see][]{buchner2014-bxa-model-comparison}, to evaluate the Bayesian evidence in support of the different models used in the fit.

We find that both a single-temperature and a two-temperature blackbody are poor fits to the X-ray data, as is the single-temperature \texttt{APEC} model. On the other hand, both a two-temperature \texttt{APEC} model and a power-law model are good fits to the data, with the latter preferred by the Bayesian analysis (the Bayes factor for the comparison of the two models is $Z_{\rm pow}/Z_{\rm 2T\_apec}\approx60$).

Figure\,\ref{fig:xspec} shows the best-fitting models for the power-law model (top panels) and two-temperature \texttt{APEC} model (lower panels). We show the pn, MOS (combined M1 \& M2), and ACIS-I spectra in the left, middle, and right panels, respectively. The thick line shows the best-fit model with the 1$\sigma$ and 3$\sigma$ range of uncertainty indicated in the shaded regions. For the two-temperature model, we also show the two individual components. The posterior distribution for the fitted parameters of the power-law and \texttt{APEC} models can be found in Appendix Fig.\,\ref{fig:X-ray-corner-apec-apec}. We quote all parameter uncertainties at the 1$\sigma$ level. For the power-law model, we find a photon index of $\Gamma=2.43^{+0.17}_{-0.15}$. For the two-temperature plasma model, we find a soft component with temperature $kT=0.22\pm0.03$\,keV, and a harder component with temperature $kT = 4.2\pm1.1$\,keV. We estimate the flux in the 0.25--10\,keV band by integrating the best-fitting power-law model across this energy range. We find a flux of $F_{\rm X}[0.25-10.0\,\rm{keV}] = 7.45_{-0.69}^{+0.68} \times 10^{-15}\,\mathrm{erg\,s^{-1}\,cm^{-2}}$ (comparable with the previous \textit{Chandra} estimates). At a measured distance of $D=41.4\,\rm{pc}$, this yields an X-ray luminosity of $L_{\rm X}[0.25-10.0\,\rm{keV}] = 1.44\pm0.13 \times 10^{27}\,\mathrm{erg\,s^{-1}}$.

To derive an easy comparison with other X-ray emitting WDs that are powered by accretion, we derive the accretion rate that would be implied by the measured X-ray luminosity if the X-rays were generated by material infalling from infinity to the surface of the WD. As we explain in more detail below, however, the rapid rotation and high magnetic field of the WD place the object firmly in the propeller regime, so if accretion is indeed powering the X-rays, the amount of infalling plasma is likely much higher.

We estimate the accretion rate from the X-ray luminosity using the expression \citep{patterson1985}

\begin{equation}
    \dot{\rm M}_X = \frac{1}{A}L_X\frac{R_{\rm WD}}{G M_{\rm{WD}}} \approx 2 \times 10^9 \,\rm{g\,s}^{-1},
\end{equation}
where for the mass and radius of the WD ($M_{\rm WD}=1.29\pm0.01\,\rm{M}_{\odot}$ and $R_{\rm WD}=2664^{+113}_{-109}\,\rm{km}$) we have used the estimates from Section \ref{sec:Modelling}.
Here, $A$ is a correction factor that accounts both for the fraction of the accretion luminosity emerging in the observed X-ray band and for geometrical effects, including emission directed away from the line of sight. Since we seek only an order-of-magnitude estimate, we set $A=1$.

\subsubsection{X-ray variability}
\label{sec:xvar}
\begin{figure}[tb]
    \centering
    \includegraphics[width=0.98\linewidth]{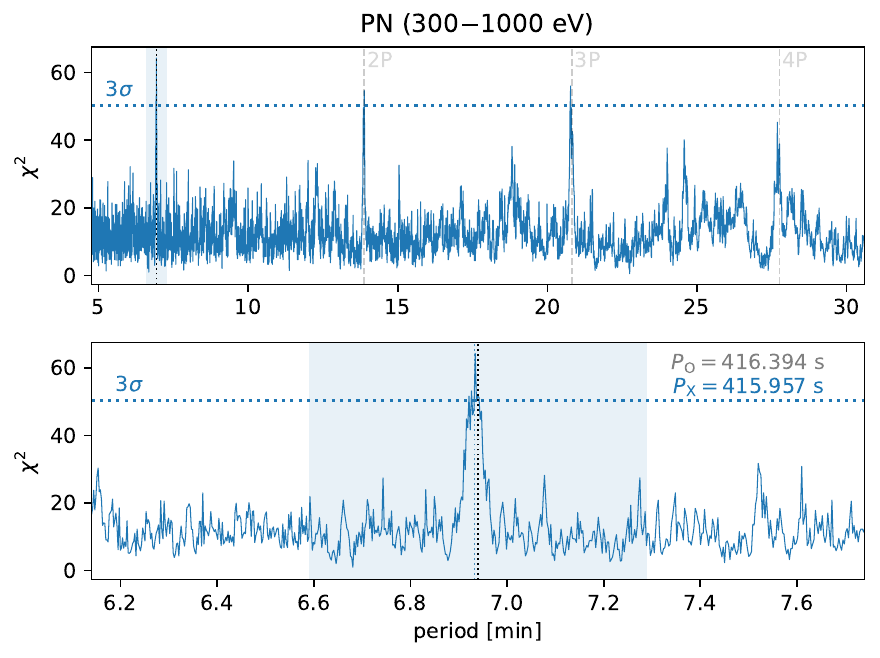}
    \includegraphics[width=0.98\linewidth]{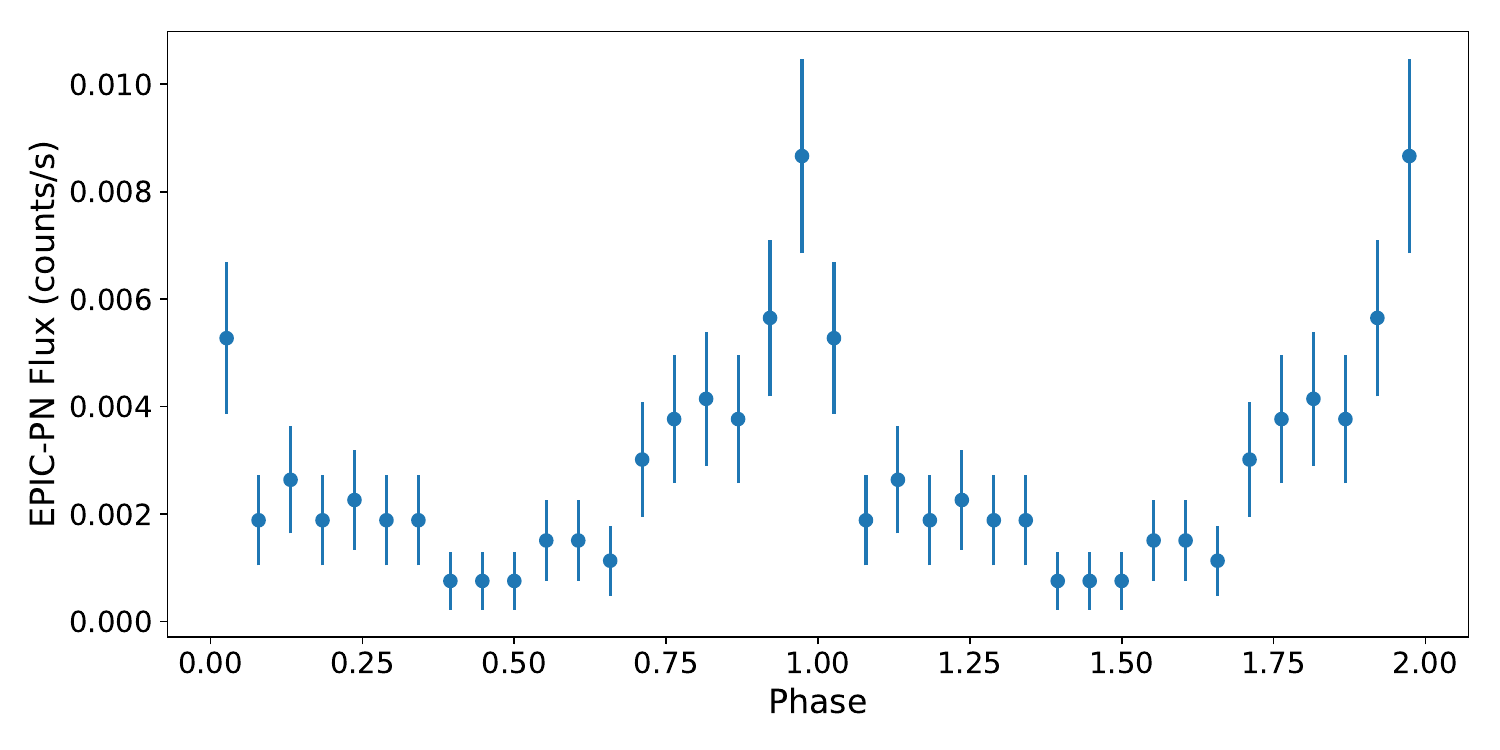}
    \caption{Top and middle: $\chi^2$ periodogram for the EPIC pn source events. The $\chi^2$ value indicates the goodness of fit for a constant count rate after phase folding the event times at each of the periods shown. Large $\chi^2$ values indicate periodic variability through statistical rejection of a constant count rate. This procedure recovers the optical photometric period. Sigma detection thresholds are defined as multiples of the standard deviation of all measured $\chi^2$ values. Bottom: Phase-folded light curve of the \textit{XMM-Newton} data.}
    \label{fig:X-ray-chisq-period}
\end{figure}

Extraction of a light curve with the SAS tools can be problematic in the low count rate, background-dominated regime because this procedure generally requires to bin event times. To investigate periodic X-ray variability, we instead phase-fold the barycentric-corrected event times onto a range of trial periods and search for evidence of periodicity through $\chi^2$ rejection of a constant count rate. We present here this analysis for events with energies in the range 0.3--1.0\,keV, and bin the counts into 12 phase bins for each trial period. This energy range is chosen to maximise the S/N. We find that our results are broadly insensitive to changes in the energy range and number of phase bins chosen.
Figure\,\ref{fig:X-ray-chisq-period} shows the results of this test on the source events measured during the observation by the EPIC pn camera. The top panel shows that the strongest period detected lies near that measured from optical photometry. The second strongest period in the top panel is a sub-harmonic, with a period of $P=2P_{\rm max}$. The middle panel shows that the period recovered in the X-ray analysis matches closely that recovered through optical photometry. The bottom panel shows the phase-folded light curve.
The light curves in the UVW1 and UVM2 filters of the Optical Monitor showed similar modulations to those obtained from COS and STIS spectra in the same wavelength ranges, but with a lower S/N.

\section{Modelling}
\label{sec:Modelling}
\subsection{New magnetic atmosphere models}
The strongly shifted Lyman components observed in the COS spectrum indicate the presence of a high magnetic field strength ($\approx$\,700\,MG) on the surface of ZTF\,J1901 and a hydrogen-dominated composition, confirming the previous estimates by C21. In the same paper, however, the characterisation of the physical properties of the WD (especially radius and effective temperature) hinged on modelling its SED with non-magnetic synthetic atmospheres. Since strong fields can significantly alter the SED of a WD, we here aim to better constrain the properties of the star by using magnetic models instead.

We have developed state-of-the-art atmosphere models that include comprehensive opacity sets \citep[e.g.][]{1974IAUS...53..265L,1992A&A...265..570J,1995A&A...298..193M,2014ApJS..212...26S,2024A&A...687A.141V} under
a variable magnetic field strength. The models are in hydrostatic equilibrium and in joint radiative and convective equilibria.
The line opacities include forbidden transitions enforced by a strong electric field \citep{2021MNRAS.507.2283Z,2022ApJS..259...47L}, particularly the 1s0\,$\rightarrow$\,3d$-$2 transitions identified in the far-UV spectra of high-field magnetic WDs.
In our models, for a given magnetic field geometry (e.g. dipole, quadrupole, multi-pole expansion) viewed from any angle \citep{1979MNRAS.189..883M,1981MNRAS.196...23M,1984MNRAS.206..407M,1989ApJ...346..444A} -- the visible
hemisphere is discretised into 900 individual surface elements characterised by a local field strength
and an angle between the field direction and the line of sight. The temperature and density
structures are computed for each surface element along with the angle-dependent intensity
spectrum. These individual spectral elements are then co-added in spherical geometry to form the
total emerging spectrum from the star. It is therefore possible to model the observed time- and angle-dependent magnetic WD spectrum for each face of the WD.

\begin{figure*}[tb]
   \centering
     \includegraphics[width=1\linewidth]{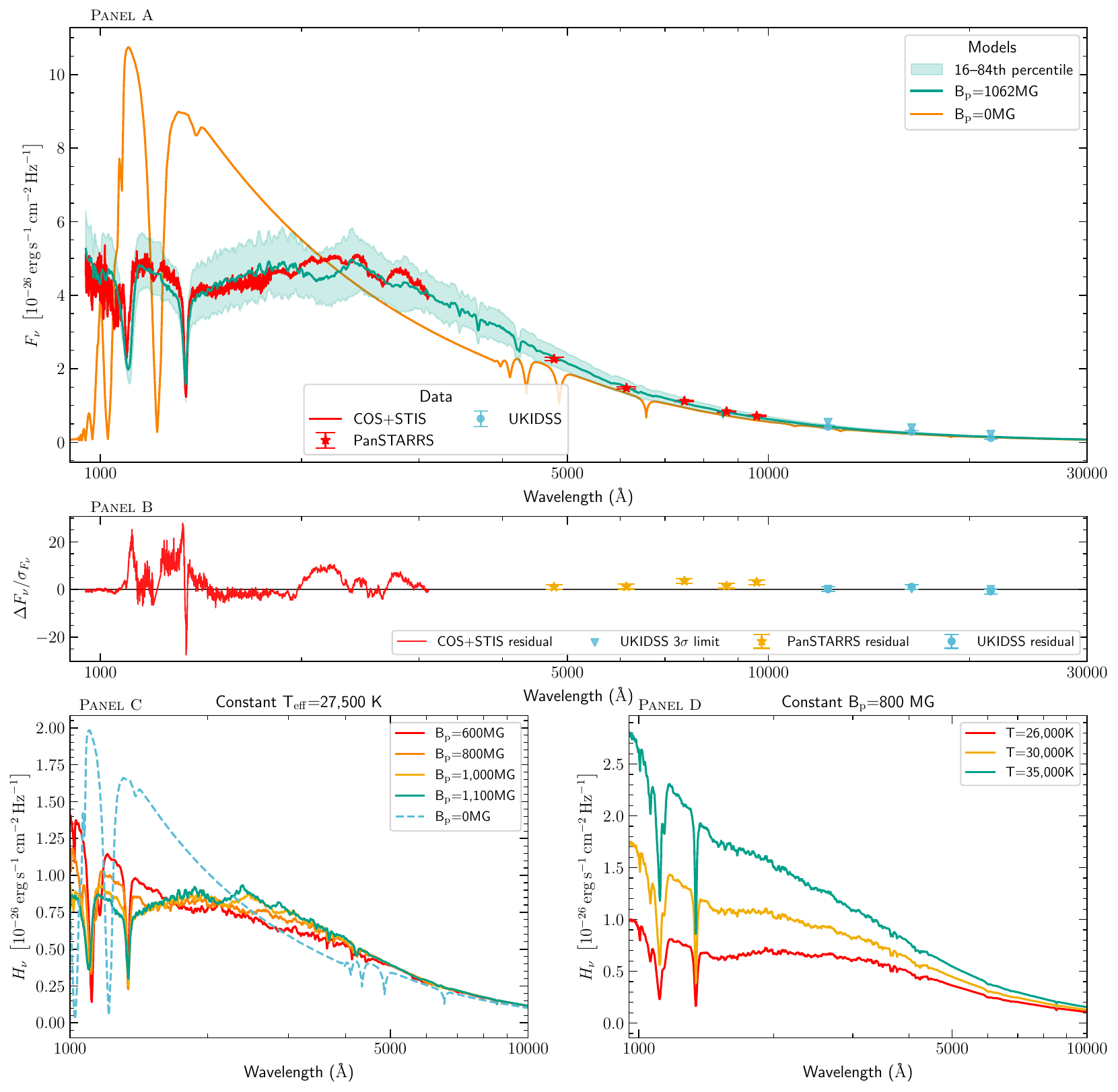}
    \caption{\textsc{Panel a}: Spectral energy distribution of J1901. The red curve shows the phase-averaged COS+STIS spectrum, the red points show Pan-STARRS photometry, and the light blue points show UKIDSS photometry. The cyan line and shaded band show the best-fitting magnetic atmosphere model and its 16--84th percentile range; the orange line shows a non-magnetic model with the same radius and temperature. \textsc{Panel b}: Residuals relative to the best-fitting magnetic model. \textsc{Panel c}: Synthetic spectra with T$_{\rm eff}=27,500$\,K, a centred dipole viewed at 80$^{\circ}$, and different polar field strengths; the dashed blue line is a zero-field model \citep{tremblay_spectroscopic_nodate}. \textsc{Panel d}: Synthetic spectra for different T$_{\rm eff}$ at fixed B$_{\rm p}=800$\,MG.}
    \label{fig:fit_model_fnu}
\end{figure*}

In \textsc{Panel c} of Fig.\,\ref{fig:fit_model_fnu}, we show several models for the same effective temperature (T$_{\rm eff}$\,=\,27,500\,K) and surface-field geometry (a dipole viewed at 80$^\circ$), but with different field strengths at the pole (B$_{\rm p}$\,=600--1,100\,MG). The dashed blue line shows a zero-field model for comparison. In \textsc{Panel d}, we keep the same field, and we change the temperature. The magnetic field changes the continuum shape, especially in the UV, by redistributing the flux across wavelength, with a stronger effect at higher fields. The field strength and structure also set the shape and location of the absorption lines, including the broad and shallow absorption features in the 2,000--3,000\,\AA\ range, which corresponds to Balmer bound-free transitions.

These new models include, for the first time, the effect on atmospheric structures of variable monochromatic opacities under a strong magnetic field. So far this effect had been neglected in the analysis of both low-field \citep[$\lesssim$\,100\,MG; e.g.][]{2009A&A...506.1341K,2013MNRAS.429.2934K,2023MNRAS.520.6111H}, and high-field magnetic WDs such as RE~J0317$-$853/EUVE~J0317$-$85.5 \citep[e.g.][]{ferrario_euve_1997,burleigh_phase-resolved_1999,vennes_multiwavelength_2003,2010A&A...524A..36K}. Although this simplifying assumption may hold true for low-field WDs, it breaks down at increasingly high fields. The new models will provide a significant improvement in the characterisation of high-field, high-temperature ($\gtrsim$ 25,000\,K) WDs compared to previous works. Improved models should eventually include exact bound-free opacities \citep{2021ApJS..254...21Z} superseding the present approximate opacities \citep[see e.g.][]{1974IAUS...53..265L}.

If a WD rotates, time-series observations, such as the phase-resolved spectroscopy that we present in this paper for J1901, can reveal the variation of the field strength and structure on the surface of the WD, as each spectrum reveals the contribution of a different region via the variation in wavelength of the absorption lines and the modulation of the shape of the SED with the rotation period. A proper modelling of these datasets, however, requires the development of an extensive suite of magnetic models with different geometries, at different magnetic field strengths and effective temperatures, and it is beyond the scope of this paper. In the next section, we employ a simple dipole structure to model the SED of the WD and constrain its physical properties.

\subsection{Modelling the spectral energy distribution of the WD}

In Fig.\,\ref{fig:fit_model_fnu}, we show the SED of the WD, combining the phase-averaged COS and STIS spectra in the UV with Pan-STARRS PS1 photometry in the optical and UKIDSS photometry in the near infrared. We employ Pan-STARRS photometry, rather than the available LRIS optical spectroscopy, because the LRIS flux calibration is not accurate enough for this purpose. The SED shows a break at $\approx\!3,000$\,\AA, where the steep optical spectrum flattens towards the UV. This behaviour has been observed in other magnetic WDs with field strengths in excess of a few hundred MG \citep{green1981,schmidt1986,gaensicke2001,cristea2026}, but was until now unexplained, with cyclotron absorption or metal absorption invoked as possible causes. In our models, the break is produced by the effect of the magnetic field on the atmospheric opacities (Fig.\,\ref{fig:fit_model_fnu}, \textsc{Panels c,d}). The best-fitting model in \textsc{Panel a}, which includes the effects of the strong field on both the monochromatic opacities and the atmospheric temperature structure, reproduces the overall SED shape.

We employ a grid of models with a dipole structure for the magnetic field with variable inclinations, field strengths, and effective temperatures to fit the SED of the WD. We use a Monte Carlo approach implemented with \texttt{emcee} \citep{foreman-mackey_emcee_2013}, leaving as free parameters the effective temperature T$_{\rm eff}$, the radius-to-distance ratio $R/D$, the dipolar field strength B$_{\rm avg}$, and the inclination of the magnetic dipole with respect to the line of sight. We derive the radius from $R/D$ using the \textit{Gaia} distance estimate ($D$\,=\,$41.44 \pm 0.08$\,pc; \citealt{gaia_collaboration_vizier_2020}).

The best-fitting model is shown in Fig.\,\ref{fig:fit_model_fnu}, and the marginalised parameter distributions are shown in Appendix Fig.\,\ref{fig:cornerplot}. The best-fit parameters are summarised in Table~\ref{tab:summary}. The revised radius of $2664^{+113}_{-109}$\,km is slightly larger than the previous estimate of $2140^{+160}_{-230}$\,km.

The best-fitting model likely captures a representative average field strength on the WD surface, because it reproduces both the SED shape and the locations of the two main Lyman absorption features. However, the shapes of the Lyman lines and of the Balmer edges in the 2000--3000\,\AA\ range are not fully reproduced, indicating that the assumed field geometry differs from the true surface field structure. As discussed above, modelling the full magnetic geometry is beyond the scope of this paper; therefore, the stellar parameters listed in Table~\ref{tab:summary} quote only statistical errors, while the simplified field model is expected to introduce larger systematic uncertainties.

\subsection{Inferring mass and cooling age}
To infer the mass and cooling age of J1901, we employed the evolutionary cooling tracks of ultra-massive WDs presented by \citet{althaus_structure_2022, althaus_carbon-oxygen_2023} for oxygen-neon (ONe) and carbon-oxygen (CO) core composition, respectively. These models provide the radius and effective temperature as a function of the cooling age of the WD for different WD masses, and incorporate detailed physics, including gravitational contraction, crystallisation, and relativistic corrections.
WDs with masses above 1.1\,M$_\odot$ that are the product of single-star evolution are expected to be composed mostly of oxygen and neon, with traces of carbon, sodium and magnesium, because the core in the progenitor star undergoes carbon burning before becoming a WD \citep{rakavy1967,murai1968,garcia-berro1997,siess2010,kippenhahn2013}, although the exact limit might depend on the progenitor star's rotation \citep{dominguez1996} and on stellar winds \citep{althaus2021}. Even in the case of merger remnants, compressional heating during the merger is expected to ignite off-centre carbon burning \citep{shen2012}, resulting in an ONe WD \citep{schwab2021evolutionary}. However, whether or not the carbon-ignition mass is different in WD mergers, and if some ultra-massive merger remnants might have CO cores, is still under debate \citep{wu2022,shen2023,sousa_optical_2023,blatman2024}. We therefore employ both models and present both mass estimates. As we derive high values for the mass, the core composition is likely ONe, no matter the evolutionary history of the WD.

We list our best estimates in Table~\ref{tab:summary}. The inferred cooling age of $\approx$\,480\,Myr is substantially older than the earlier estimate of 10--100\,Myr, which matters when constraining the formation rate of such systems.

\subsection{Cyclotron harmonics and expected signatures\label{sec:cyclotron}}
\begin{figure}[tb]
    \centering
    \includegraphics[width=0.98\linewidth]{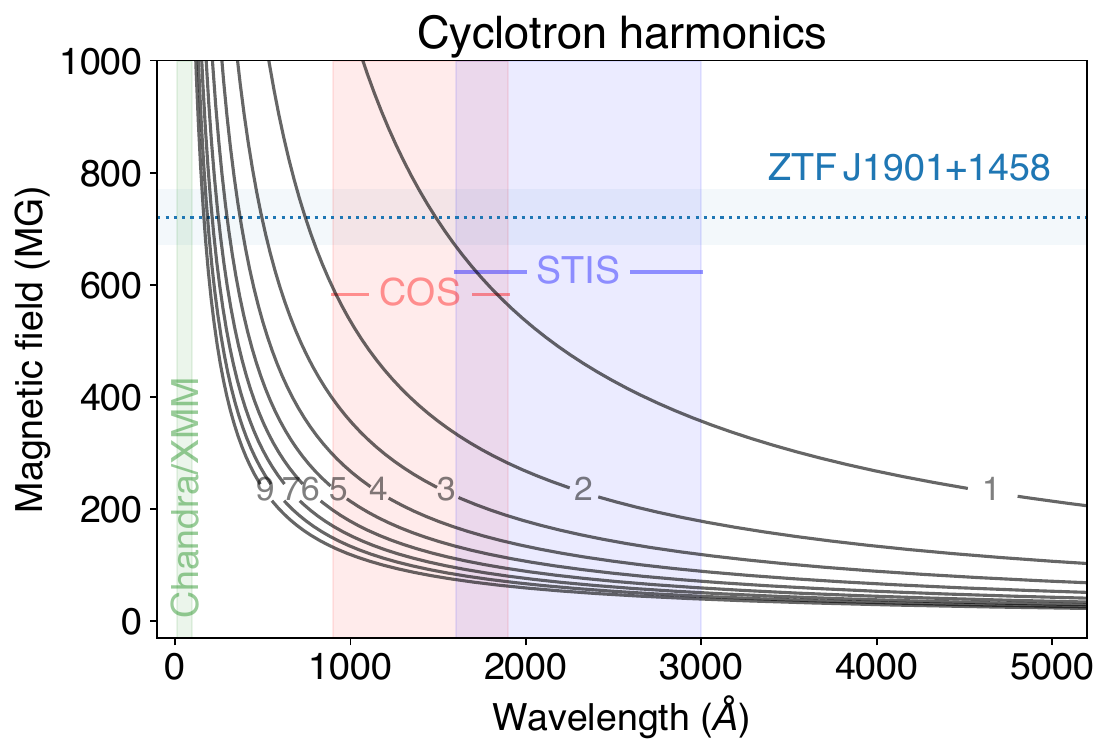}
    \caption{Central wavelength of the first nine cyclotron harmonics for mean surface magnetic field strengths. For the inferred magnetic field strength of J1901 (B$_{\rm avg}$\,$\approx$\,720\,MG), all harmonics are predicted to fall in the UV and EUV.}
    \label{fig:cyclotron}
\end{figure}

In highly magnetised WDs, cyclotron emission can appear as discrete humps in optical or UV spectra, depending on the field strength and viewing geometry. Fig.\,\ref{fig:cyclotron} plots the central wavelengths of the first nine harmonics for B$_{\rm avg}$\,=7$\times10^8$\,G. Except for the first harmonic, which overlaps with our COS spectrum, the harmonics lie in the EUV. We detect no discrete UV cyclotron feature consistent with the first harmonic. This lack of cyclotron emission in the COS spectrum might indicate that there is little active accretion onto the WD surface, or that most of the power is emitted in higher harmonics outside the observed band.

\subsection{Lack of a companion}
\label{sec:BD}

\begin{figure*}[tb]
      \centering
        \includegraphics[width=0.51\linewidth,keepaspectratio]{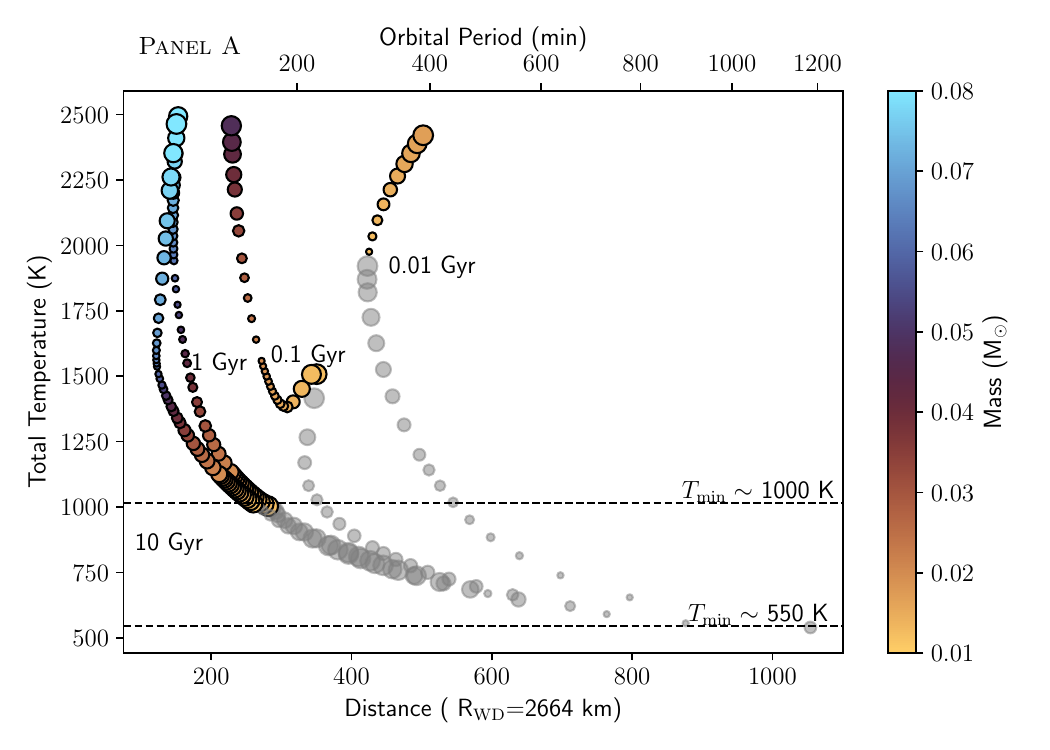}
      \includegraphics[width=0.46\linewidth]{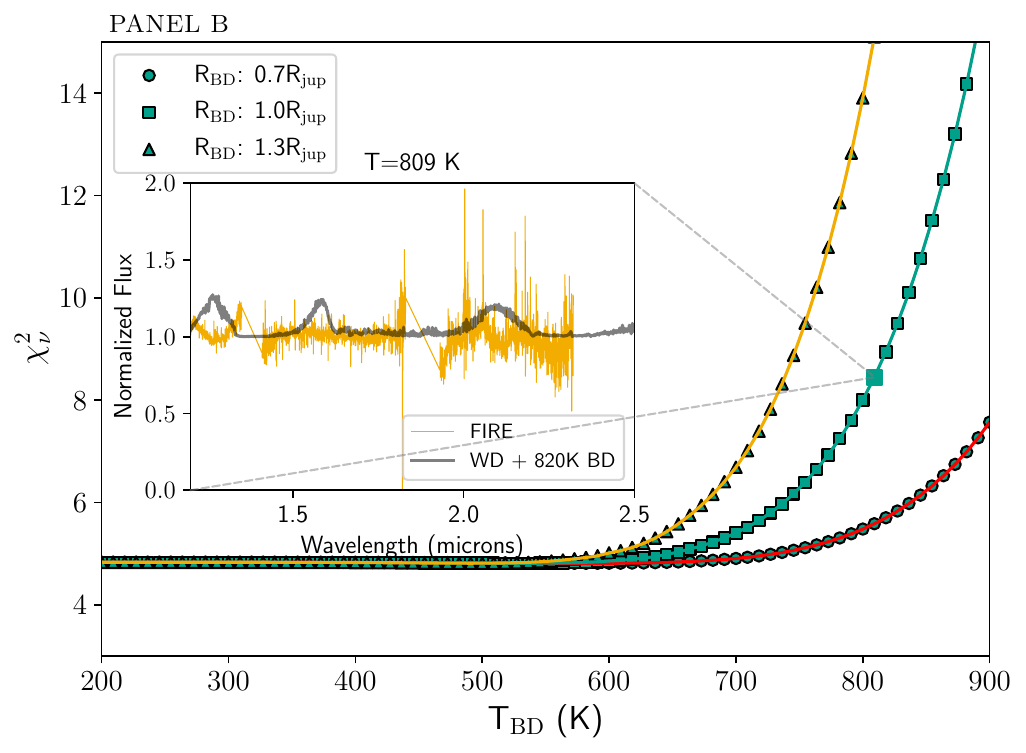}

   \caption{\textsc{Panel a}: Brown dwarf effective temperature T$_{\rm tot,BD}$ (Eq.~\eqref{eq:tot_BD}) for a near Roche-lobe-filling companion as a function of orbital separation (bottom axis, in units of the WD radius, R$_{\rm WD}$\,$\approx$\,2664\,km) and corresponding orbital period (top axis). The coloured points denote different brown dwarf masses (colour bar in M$_\odot$). The grey points mark companions with $M\leq0.01$\,M$_{\odot}$. The marker size scales with brown dwarf radius, and the ages are labelled beside the corresponding tracks.
   \textsc{Panel b}: Reduced $\chi^{2}_{\nu}$ for our fitting of a WD+brown dwarf model to the FIRE spectrum as a function of brown dwarf temperature. The different markers correspond to tracks of constant brown dwarf radius (R$_{\rm BD}$) for fixed $\log(g)=4$.}
   \label{fig:BD_M-R-Relation}
\end{figure*}
A common source of X-rays in WDs is accretion from a stellar or substellar companion. We here analyse the possibility that J1901 is the accretor in a binary system. As shown in Fig.\,\ref{fig:fit_model_fnu}, \textsc{Panel a}, the near-infrared flux at $1$--$2.5\,\mu\mathrm{m}$ is consistent with the WD continuum alone: the best-fitting magnetic model, which was fit only to the optical and UV data, passes through the UKIDSS photometry without showing any excess. The lack of an excess in the optical or infrared, as well as the absence of molecular absorption features in the optical spectrum, excludes the presence of a low-mass star, leaving only a brown dwarf as a possible companion. Because the WD is hot, any brown dwarf close enough to partially fill its Roche lobe would be irradiated. We estimate the irradiation temperature as \citep{jermyn_tidal_2017,lothringer_atmosphere_2020}
\begin{equation}
    \rm{T}_{\rm irr,BD} \approx \rm{T}_{\rm eff,WD}\sqrt{\frac{R_{\rm WD}}{2a}}\sqrt[4]{(1-A_B)}
    \label{BD_irr},
\end{equation}
where $A_B$ is the Bond albedo of the brown dwarf, $R_{\rm WD}$ and T$_{\rm eff,WD}$ are the WD radius and effective temperature, and $a$ is the orbital separation. We assume the flux on the surface of the brown dwarf to be the sum of the internal flux and the irradiated flux, and therefore we take the effective temperature to be
\begin{equation}
    \rm{T}_{\rm tot,BD} = \sqrt[4]{\rm{T}_{\rm irr,BD}^4 + \rm{T}_{\rm int, BD}^4}\,,
    \label{eq:tot_BD}
\end{equation}
where T$_{\rm int, BD}$ would be the effective temperature of an isolated brown dwarf at its current evolutionary state, which we estimate using the brown dwarf evolutionary and spectral models of \citet{marley_sonora_2021}. More massive brown dwarfs fill their Roche lobes at smaller separations, where they intercept a larger fraction of the WD luminosity. For example, a 0.05\,M$_{\odot}$ brown dwarf at 1\,Gyr would fill its Roche lobe at $\approx$\,170\,R$_{\rm WD}$ (orbital period $P\sim$\,1.1\,h) and reach T$_{\rm irr, BD}\sim$\,2000\,K from Eq.~\eqref{BD_irr}. In contrast, a 0.01\,M$_{\odot}$ brown dwarf at 1\,Gyr must orbit farther out ($\sim$\,1\,R$_\odot$, $P\sim$\,2.7\,h) and would be heated to only T$_{\rm irr}\sim$\,1000\,K. We conservatively assume a moderately high albedo \citep[$A_B=0.5$; brown dwarfs are expected to have lower albedos,][]{marley1999,hernandez-santisteban2016} and full heat redistribution, so that T$_{\rm irr}$ represents the roughly uniform day-side surface temperature. \textsc{Panel a} of Fig.~\ref{fig:BD_M-R-Relation} shows the brown dwarf temperature as a function of separation, assuming near Roche-lobe-filling. The upper x-axis shows the corresponding orbital period. Tracks show different model ages, marker size scales with radius, and the colour bar shows the brown dwarf mass. Planetary-regime models ($<$\,13\,M$_{\rm Jup}$) are shown in grey. The minimum brown dwarf irradiation temperature attainable is T$_{\rm tot,BD}\approx$\,1000\,K for a 1--10\,Gyr brown dwarf. We also mark the lowest-mass planet in the Sonora grid ($\approx$\,0.5\,M$_{\rm Jup}$), which would fill its Roche lobe at a period of $\approx$\,1100\,minutes with a temperature of $\approx$\,550\,K.

The FIRE spectrum allows us to exclude even colder objects. We cannot use its absolute flux directly because of known FIRE flux-calibration limitations \citep[see][]{simcoe2008,sullivan_calibrated_2012}; for example, \citet{vito_chandra_2021} report a FIRE normalisation $\approx$\,15\% lower than previous spectra of the same object, attributing the difference to seeing. However, cool brown dwarfs have strong molecular absorption features, so the lack of such features in the smooth FIRE spectrum constrains the temperature of a possible companion. We compare the observed FIRE spectrum with combined WD+brown dwarf model spectra, using the best-fitting magnetic WD atmosphere and brown dwarf spectra from \citet{marley_sonora_2021}. Since the absolute FIRE flux calibration is unreliable, we normalise both the observed and model spectra: the FIRE spectrum is normalised with a 4th-order polynomial, and the combined model is normalised by the WD continuum. We then compute $\chi^2_\nu$ as a function of brown dwarf temperature (Fig.\,\ref{fig:BD_M-R-Relation}, \textsc{Panel b}), assuming fixed $\log(g)=4$ and radii between 0.7 and 1.3\,R$_{\rm Jup}$. At low temperatures, the model absorption features are below the FIRE sensitivity, and $\chi^2_\nu$ remains near $\approx$\,4.8, reflecting correlated calibration noise larger than the formal uncertainties. At higher temperatures, the molecular features become detectable and $\chi^2_\nu$ rises sharply. We therefore exclude brown dwarfs with temperatures above approximately 700\,K.

\section{Discussion \label{sec:Discussion}}

\subsection{An extreme white dwarf merger remnant}
In the previous section, we used our new magnetic models and the new \textit{HST} spectroscopy to derive stronger constraints on the physical parameters (temperature, radius, mass, cooling age and magnetic field strength) of J1901. The updated parameters, with the exception of the effective temperature, are broadly consistent with those reported in C21, and confirm the nature of the WD as an extremely massive, highly magnetised and rapidly rotating merger remnant.
The slightly larger radius that we derive (by a few hundred kilometres) and correspondingly lower mass do not alter the basic picture: this star is one of the most compact and massive merger remnants known. However, the new radius and mass estimate change the expected central density of the WD. Using the density profiles from \citet{althaus_structure_2022,althaus_carbon-oxygen_2023}, we find the central density of the WD to be $7.55 \times 10^8$ g cm$^{-3}$ and $7.31 \times 10^8$ g cm$^{-3}$ for the CO and ONe core compositions, respectively. These estimates are close to but below the threshold for electron capture onto sodium \citep[$1.7 \times 10^9$\,g\,cm$^{-3}$,][]{schwab_importance_2017}. Moreover, \citet{althaus_structure_2022} predict that an ONe core WD at this mass and temperature should already have crystallised $\approx$\,50\,\% of its stellar mass ($\approx$\,80\,\% in the case of a CO core). It is therefore unlikely that the star is undergoing Urca cooling at this moment, as previously suggested by C21 and \citet{schwab_cooling_2021}. However, J1901 could still undergo important $^{22}$Ne distillation during crystallisation. As discussed by \citet{salaris_ne22_2024}, the solid phase that forms within the core is relatively depleted in $^{22}$Ne, which remains in the surrounding liquid and can migrate inward under gravity. This process releases additional energy and can delay cooling substantially, potentially affecting the evolutionary timescales for a star with a large crystallised fraction such as J1901.
In the models, the central temperatures in the core, which are important to constrain axion production, are $\approx$\,1.45$\times 10^{7}$\,K for both the CO and ONe core compositions.

We note that, since the errors on the effective temperature and radius are underestimated, we assume an average magnetic field geometry, as opposed to fitting for the geometry using the phase-resolved spectra. Thus, we are also underestimating the errors on the mass and cooling age. Since the central density estimate is strongly dependent on the mass value that we derive from the SED fitting, our estimate for electron capture is open to revision. We intend to address this factor in a follow-up study on J1901.

\begin{figure}[tb]
   \centering
   \includegraphics[width=1\linewidth]{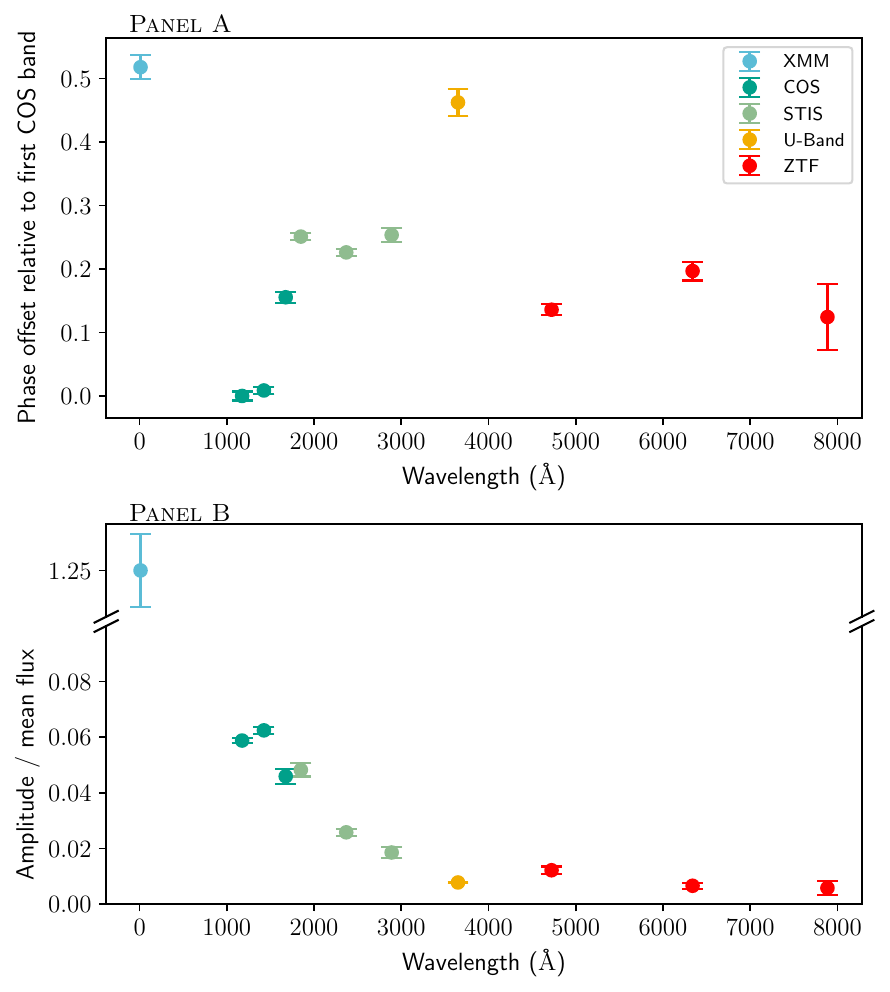}
   \caption{\textsc{Panel a}: Phase of the minimum in the sinusoidal light curve fit in the different bands, measured relative to the bluest COS band (C1). This reference point highlights the near anti-phase relation between the X-ray light curve and the bluer UV bands. \textsc{Panel b}: Amplitude of the light curve in the same bands.}
   \label{fig:lc-comp-normamp}
\end{figure}

\subsection{Origin of the optical and UV variability}
\label{sec:UVOrigin}
\begin{figure*}[h]
    \centering
    \includegraphics[width=0.9\linewidth]{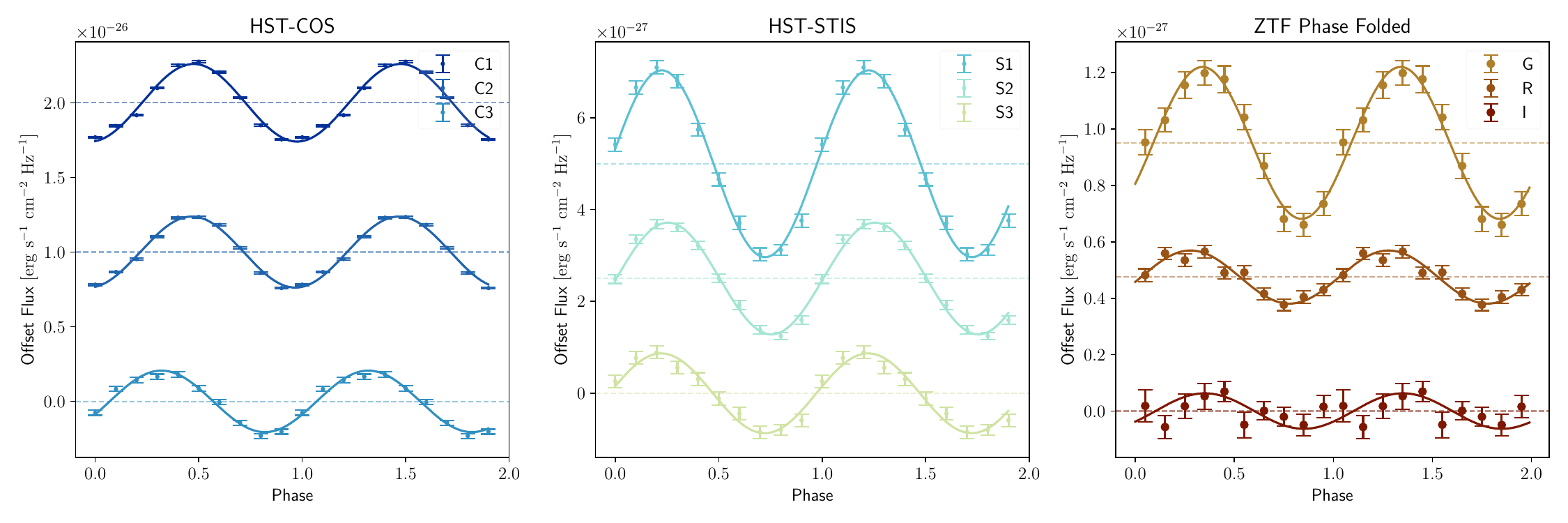}
    \includegraphics[width=0.9\linewidth]{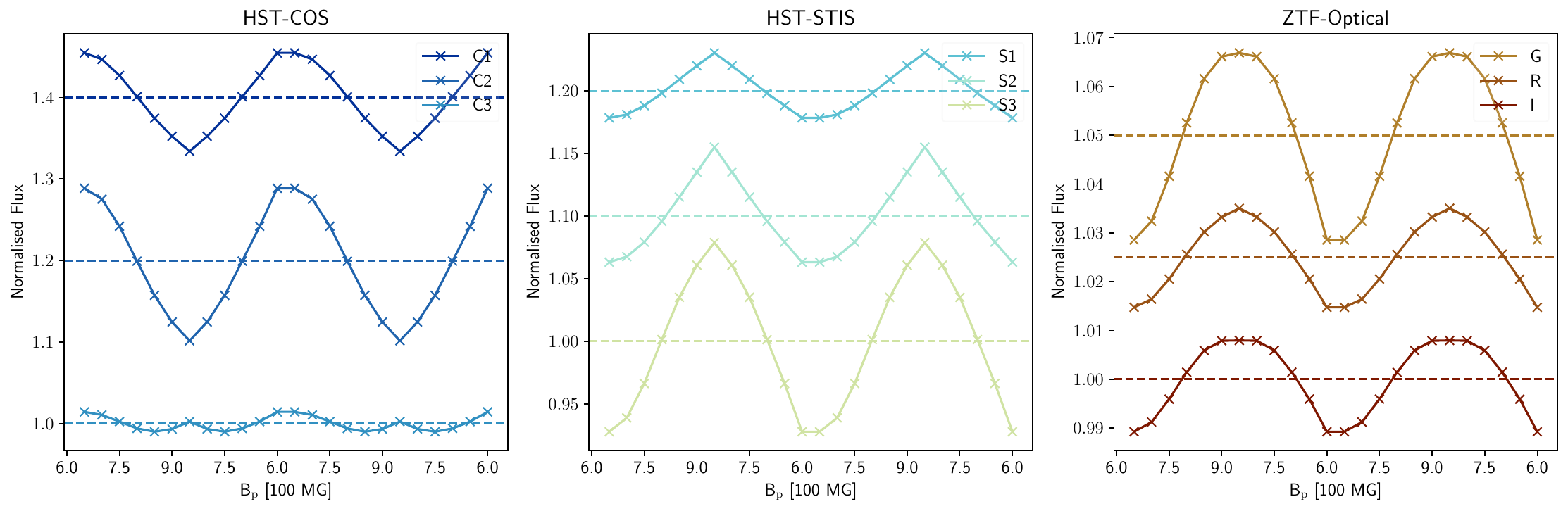}
    \caption{\textsc{Panels a,b,c}: Phase-folded and normalised light curves in the different bands. The phase is defined with respect to the reference ephemeris. The ZTF light curves are in \textsc{Panel c}, while the \textit{HST} light curves for COS and STIS are in Panels a and b, respectively. \textsc{Panels d,e,f}: Flux variation as a function of magnetic field strength constructed for the same spectral bands using magnetic atmosphere models with constant $T_{\rm eff}\approx27,500$\,K (Fig.\,\ref{fig:fit_model_fnu}). The flux is normalised and offset in each band. These curves are not quantitative light curve models because they use a static geometry, but they illustrate the wavelength-dependent effect of magnetic dichroism.}
    \label{fig:stephane-models-lc}
\end{figure*}

As shown in Fig.\,\ref{fig:ly-alpha-fit}, the Lyman absorption features vary in centroid and width over the rotation period, indicating a change in the visible magnetic field strength across the WD surface. The continuum across the UV and optical also varies with period, showing a sinusoidal modulation whose phase depends on wavelength. To study the UV variability, we extracted \textit{HST} light curves in six ad hoc bands: three in the COS wavelength range (C1, C2, and C3, with central wavelengths 1175, 1424, and 1674\,\AA) and three in the STIS wavelength range (S1, S2, and S3, with central wavelengths 1848, 2368, and 2889\,\AA). Folded on the WD spin period, these light curves show an approximately sinusoidal variation with increasing amplitude towards bluer wavelengths (see \textsc{Panels a, b} of Fig.\,\ref{fig:stephane-models-lc}). Figure~\ref{fig:lc-comp-normamp} compares the fitted UV phases and amplitudes with those from ZTF $g$, $r$, and $i$ and from LightSpeed $u$: the amplitude increases monotonically from the red optical to the far UV, while the phase shift is not monotonic. We also show the amplitude and phase of the X-ray light curve from \textit{XMM-Newton}.

The variability in the continuum is most naturally explained by the changing magnetic field visible on the WD surface \citep{ferrario_magnetic_1997}. As the WD rotates, regions with different field orientations and local strengths contribute to the emission. As a proof of concept, we take magnetic models with the same effective temperature (27,500\,K) and field geometry (a dipole viewed at 80$^\circ$), but with field strengths between 600 and 900\,MG (Fig.\,\ref{fig:fit_model_fnu}, \textsc{Panel c}), and extract the flux in the observed bands ($g$, $r$, and $i$ in the optical, and C1--S3 in the UV). The lower panels of Fig.\,\ref{fig:stephane-models-lc} show the flux variation between these models as a function of magnetic field strength; each marker is a model, and we vary the field up and down between 600 and 900\,MG twice. These curves are qualitatively similar in amplitude to the observed light curves and display phase shifts between bands, although not exactly the same shifts as in the observed light curves. The curves in \textsc{Panels d, e, f} are not realistic models for the observed light curves because the geometry of the magnetic field is kept fixed, and only the field strength varies, while in a rotating WD, the visible field orientation would also change. However, this proof of concept shows that the observed variability can, in principle, be explained by the changing surface field strength without invoking temperature variations, although spots, temperature variations, and composition inhomogeneities across the surface could also contribute to the observed variability. A full phase-resolved analysis, including synthetic spectra with varying magnetic field inclination, is needed to refine these estimates and will be the focus of a follow-up work.

\begin{figure*}[tb]
   \centering
    \includegraphics[width=2\linewidth,height=0.46\linewidth,keepaspectratio]{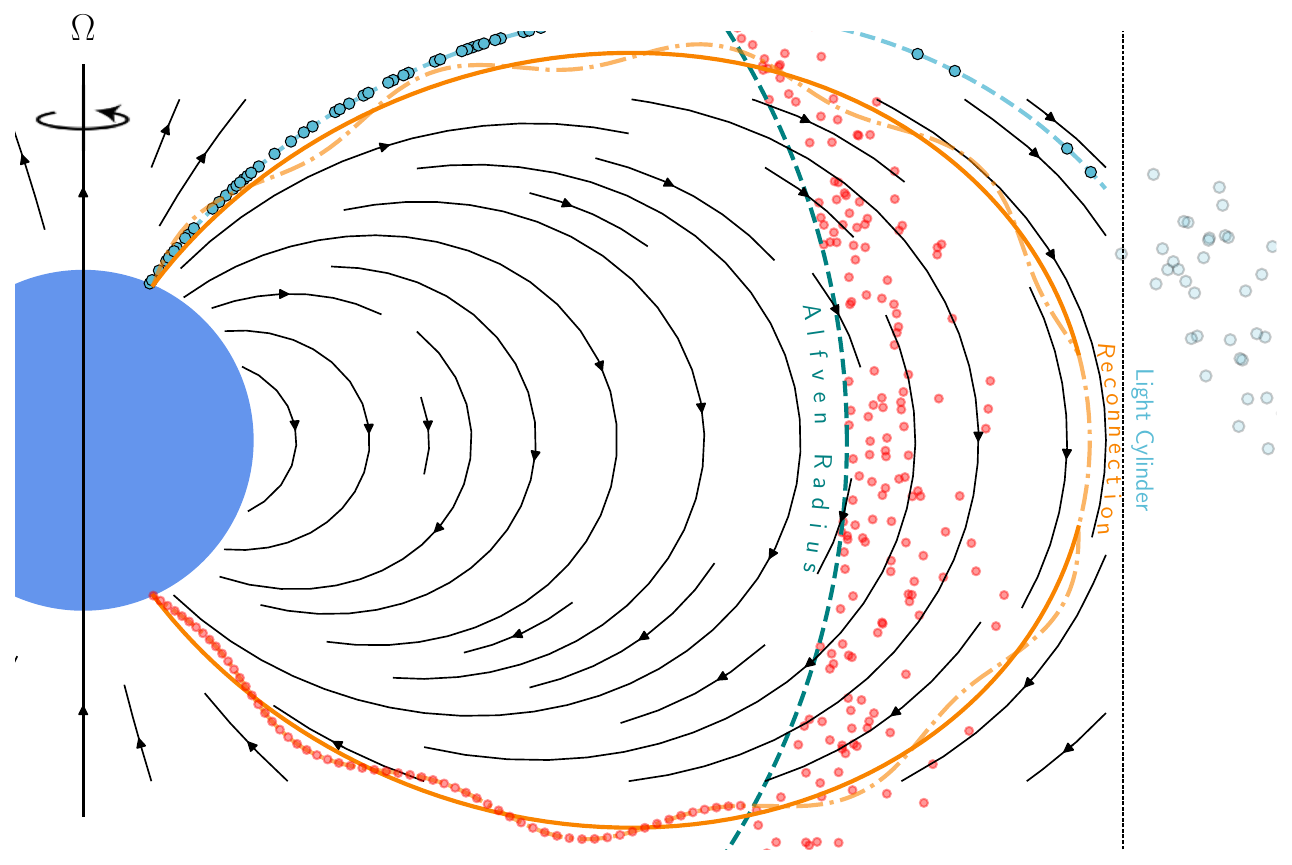}
    \caption{Schematic, not to scale, of one possible configuration for J1901. The WD is shown in blue. For simplicity, the magnetic field is shown as a dipole aligned with the rotation axis. As we see variations in the spectra with phase, we know that the magnetic moment cannot be aligned with the rotation axis. The Alfv\'{e}n and light cylinder radii are marked with dashed lines. The red particles show material infalling from outside the magnetosphere;
    once it crosses the Alfv\'{e}n radius, material is forced to follow magnetic field lines. If the Alfv\'{e}n radius lies outside of the corotation radius, most of the infalling material would be ejected from the system; shocks or reconnections in this ejection region could be the source of the observed X-rays. Alternatively, the pressure exerted by the infalling material on the edge of the magnetosphere could cause Alfv\'{e}n waves to propagate through the magnetosphere, causing reconnections close to the surface (orange line).
    The blue particles illustrate instead the possibility of an outflow of material extracted by the potential difference across open field lines at the light cylinder. The Alfv\'{e}n radius shown corresponds to a higher accretion rate than the rate inferred from the X-rays: at the inferred rate, the Alfv\'{e}n radius would lie outside the light cylinder. A large reservoir of material around J1901 is required to push the Alfv\'{e}n radius within the light cylinder.}
   \label{fig:Illustration}
\end{figure*}

\subsection{Origin of the X-ray emission}
\label{sec:origin}
Hot WDs (T$_{\rm eff}$\,$\geq$\,25,000\,K) can be detected in the X-rays because of their soft photospheric emission \citep[see e.g.][]{jordan1994,vennes1999,odwyer2003}; however, the X-ray spectrum of J1901 is too hard and too bright to be consistent with photospheric emission:
The observed X-rays must therefore be powered by a different mechanism, most likely by the interaction of the WD or its magnetosphere with circumstellar material. From the infrared analysis, we exclude the possibility of the WD being the accretor in a mass-transferring binary with a close to Roche-lobe-filling stellar or substellar companion (see Section~\ref{sec:BD}). The coldest brown dwarf that would be consistent with the infrared data has a temperature below 700\,K (see Fig.\,\ref{fig:BD_M-R-Relation}); this is significantly below the lower limits on irradiation temperatures for any brown dwarf above 13\,M$_{\rm jup}$.

It would still be possible for a cold planet to be the mass donor, because it would fill its Roche lobe at a larger orbital period, or for a more distant brown dwarf that does not fill its Roche lobe to transfer material through a weak wind. Both scenarios are unlikely: the presence of a close-in planet in a double-degenerate merger system would be difficult to explain, and no direct evidence of winds has yet been observed from any brown dwarf. We therefore explore other explanations for the origin of the circumstellar material. It is already known that accretion of planetary material onto the surface of a WD can power the emission of soft X-rays: X-rays have been detected from the prototypical polluted WD G\,29--38 \citep{cunningham2022}, and the accretion rate needed to power the X-ray emission in G\,29--38 is very similar to the one needed for J1901 (\,$\approx$\,$10^9$g\,s$^{-1}$). Over a quarter of WDs show evidence of accretion of planetary material as their optical and UV spectra show the presence of metals in their atmosphere \citep[called polluted or planetary-enriched WDs;][]{koester14,zuckerman03,zuckerman10,2024ApJ...976..156O}. Due to their high surface gravity, WD spectra usually only show the lightest element present (most commonly hydrogen or helium), thus, the presence of metals in the spectra of polluted WDs indicates recent accretion of debris from planetary bodies \citep[e.g.][]{zuckerman07,xu17,vanderburg2015,gaensicke2019,manser2019}.

In the case of J1901, if a populated planetary system surrounded the original stellar binary, it has now undergone several disrupting events, possibly including one or two common-envelope events, two AGB phases and a double-WD merger. No studies have been performed to analyse the combined effects of such events on the dynamics of a planetary system, and it is not clear if planetary objects can survive through to the final merger remnant stage. Contrary to G\,29--38, J1901 does not show any prominent metal absorption features in its spectrum, and its SED does not present any infrared excess, indicating the presence of a dusty debris disc. Neither excludes the planetary accretion scenario, however, as metal lines in the spectrum could be easily washed away by magnetic broadening, and we do not expect a debris disc to be close enough to the WD to emit in the near infrared because it would be disrupted by the strong magnetic field. Future observations (for example, deep mid-infrared imaging or spectroscopy with the JWST) could detect the faint signature of a disc at larger distances from the WD.

As J1901 is most likely a WD merger remnant, a possible origin of circumstellar material would be the fallback of bound material ejected during the merger event. Simulations of double WD mergers predict that right after the merger, the remnant WD is surrounded by a thick disc and an extended tidal tail \citep{guerrero2004,loren-aguilar_high-resolution_2009,dan2014structure}. The material in the disc is expected to be accreted on a viscous timescale, while material at high eccentricity in the bound tails accretes at longer timescales, with the fallback accretion rate predicted to decline as a power law, roughly as \(\dot{\rm M} \propto t^{-5/3}\) \citep[see e.g][]{rosswog2007,loren-aguilar_high-resolution_2009,ishizaki_fallback_2021}.
For a system with a cooling age of several hundred Myr, the corresponding accretion rate is extremely low, consistent with our X-ray luminosity estimates. However, most available simulations have only analysed the first few minutes or days after the merger, and no studies have been performed to explore if a very low level of accretion could still be ongoing hundreds of millions of years after the merger.

Bondi accretion from the local ISM could be another source of material, given the faintness of the X-rays. However, J1901 is only $\approx$\,40\,pc from Earth, so it lies in the Local Bubble, a region surrounding the Solar system where the ISM densities are very low \citep[$\sim$\,0.1 cm$^{-3}$, see][]{cox_local_1987,frisch_interstellar_2011}. Using the expression for Bondi accretion rate from \citet{wesemael_accretion_1979}, assuming nominal WD velocities in the local neighbourhood ($\approx$\,30\,km\,s$^{-1}$, consistent with the Gaia proper motion of J1901), we obtain an accretion rate of $\sim$\,$2\times10^9$\,g\,s$^{-1}$ from the ISM. If we convert the observed X-ray luminosity of J1901 into an accretion rate by assuming that half of the gravitational energy of the infalling material is emitted in X-rays that we can detect \citep{patterson1985}, we find a lower limit on the accretion rate of $\dot{\rm M}\sim$\,$10^9$\,g\,s$^{-1}$. Since we expect most of the infalling material to be ejected in a propeller mechanism \cite{wesemaelAccretionGrainsElement1982,eracleous1996}, it is unlikely that the ISM could provide the reservoir of material needed to power the X-ray emission. Furthermore, we have shown that in the twin system of J1901, J2008, accretion from the ISM could not explain the observed X-rays nor the large spin derivative \citep{cristea2026}, so we exclude the ISM as a likely explanation for the X-ray emission in J1901 as well.

One possible origin of circumstellar material could be the WD itself, if some ionised gas could be extracted from its surface. Because of the low temperature and extreme surface gravity of the WD, a radiative or line-driven wind would be impossible \citep[see e.g.][]{vennes1988,unglaub2008}; however, the rapidly rotating magnetic field could potentially drive a small outflow. The footprint on the surface of the WD of the open field lines that reach the light cylinder\footnote{The light cylinder is defined as the distance from the rotation axis of the WD at which the corotation velocity becomes equal to the speed of light. Magnetic field lines that reach the light cylinder cannot close back to the surface of the WD.} is not negligible \citep[for a dipolar structure, the polar cap radius would be about 30\,km, following][]{goldreich_pulsar_1969}.
Open field lines in a vacuum create a potential difference between the surface and the light cylinder, which, in the case of J1901 would be of the order of $\approx$\,10$^{11}$V.
We note that in a previous study on J1901 and other highly magnetised WDs \citep{bamba_x-ray_2024}, it has been suggested that the X-ray emission in these objects is due to curvature radiation from electrons accelerated to relativistic speeds by this potential difference along magnetic field lines, similar to the mechanism powering the X-ray emission in pulsars. In the case of pulsars, however, a gap is formed outside the neutron star surface because it is hard to extract ions from the degenerate and crystallised crust, and relativistic currents are thought to be generated when the vacuum is broken, and electron-positron pairs are created \citep[see e.g.][]{ruderman_theory_1975}. The surface of a WD, on the other hand, is made of partly ionised gas, from which it would not be possible to extract only electrons, as ions would follow. The potential difference is likely neutralised by a slow, magnetically driven wind.
Available studies on magnetically driven winds are mainly focused on main sequence and evolved stars and mostly on how the field accelerates winds that are powered by other mechanisms, as radiation or reconnections \citep{weber_angular_1967,belcher_magnetic_1976,goldreich_stellar_1970,thirumalai_hybrid_2010,vidotto_stellar_2014,johnstone_fast_2017}. No studies are available that examine the steady state of a wind in the case of an extreme and old object such as J1901 (for the classical \citealt{weber_angular_1967} model, the three critical points in the solution lie outside the light cylinder for J1901). Estimating whether the rapidly rotating field can sustain a significant outflow from the surface of the WD, this late into its evolution, would require a self-consistent MHD simulation, which is beyond the scope of this paper. Recent MHD studies aimed at modelling IRAS 00500+6713, the stellar remnant of the historical supernova 1181 \citep{gvaramadze2019,oskinova_x-rays_2020,ritter2021,schaefer2023,lykou2023,fesen2023,cunningham2024}, show that strong winds can be sustained in the early post-merger stages, when the remnant WD is extremely hot and nuclear burning might still be dominating the energy budget \citep[][]{kashiyama_optically_2019,zhong_optically_2023,ko_radio_2024}. However, if we assume that a small outflow can be extracted and accelerated by the field at the expense of the rotational energy of the WD, this could explain the observed X-rays, as shocks could arise within the wind, or the pressure of the wind on the magnetosphere could lead to reconnections (we analyse possible emission mechanisms in the next section).

\subsection{X-ray emission mechanism}
\label{sec:x-ray-mechanism}

We have argued that the X-rays from J1901 are likely produced by the interaction of the WD, or its magnetosphere, with a reservoir of circumstellar plasma. This material could either be supplied externally and interact with the magnetosphere as it moves inward, or be extracted from the WD itself in a magnetically driven wind. We now consider where the emission may arise and which physical mechanisms are most plausible.

Because of the extreme magnetic field and the rapid rotation of J1901, we expect that any infalling material towards the WD would be mostly ejected in a propeller mechanism, as is seen in WD propellers such as AE Aquarii and in transitional pulsars \citep[see e.g.][]{wynn1997,eracleous1996,campana2018,papitto2022}. In fact, the high field forces ionised gas to follow magnetic field lines in the region close to the WD, within the magnetospheric radius, which is usually taken as a fraction ($\xi$) of the Alfv\'{e}n radius \citep{pringle1972,ghosh1979,papitto2022}

\begin{align}
    \rm{R}_{\rm m} &\approx \xi \rm{R}_{\rm A} \approx \xi \left(\frac{\rm{B_p}^2 \rm{R}^6_{\rm WD}}{\dot{M}\times\sqrt{2GM_{\rm WD}}}\right)^{\frac{2}{7}} \nonumber\\
    &\approx 700~ R_{\rm WD} \left(\frac{\xi}{0.5} \right) \left(\frac{\rm{B_p}}{800\,\rm{MG}}\right)^{\frac{4}{7}} \left(\frac{\rm{R_{WD}}}{2600\,\rm{km}}\right)^{\frac{12}{7}} \nonumber\\
    & ~~~~~~~~~~~~~~~~~~~~~~~~~ \times \left(\frac{\rm{M_{\rm WD}}}{1.3\,\rm{M}_\odot}\right)^{-\frac{1}{7}} \left(\frac{\dot{\rm M}}{10^{16}\,\rm{g\,s}^{-1}}\right)^{-\frac{2}{7}}
    \label{eq:ra}
\end{align}
where $B_p$ is the magnetic field at the surface and $\dot{\rm M}$ is the mass infall rate at the magnetospheric radius. Even at a relatively high infall rate of $10^{16}$\, g\,s$^{-1}$, the magnetospheric radius is seven hundred WD radii.
At these large radii, the centrifugal acceleration of corotating material is much larger than the centripetal attraction due to gravity. These two forces become equal at the Keplerian corotation radius:

\begin{equation}
    \rm{R_{K}} = \left(\frac{G\rm{M_{WD}}}{\Omega^2}\right)^{\frac{1}{3}} \approx 35~ \rm{R_{WD}} \left(\frac{P}{6.9\,\rm{min}}\right)^{\frac{2}{3}} \left(\frac{M_{\rm WD}}{1.3\,\rm{M}_\odot}\right)^{\frac{1}{3}} \, .
    \label{eq:rk}
\end{equation}

\noindent
Unless the accretion rate is extremely high, in excess of $10^{20}$\,g\,s$^{-1}$, the magnetospheric radius is expected to be larger than the corotation radius, placing J1901 in the propeller regime: most ionised gas that penetrates the extended magnetosphere is centrifugally expelled long before it can reach the stellar surface.

Simulations of fast-rotator propellers show that even when $r_{\rm A}\!\gg\!r_{\rm co}$, however, interchange instabilities and field-line inflation can produce transient funnel streams that allow some material to penetrate the centrifugal barrier \citep{romanova_propeller-driven_2005}. If the X-rays that we measure are powered by accretion, then the small accretion rate that we estimate in Section~\ref{sec:xrayspec} ($\approx\!10^{9}\,\mathrm{g\,s^{-1}}$) could be explained by a small percentage of infalling material being channelled along magnetic field lines and producing compact hot spots near the magnetic poles, while most of the material infalling would be ejected at the magnetospheric radius. Depending on the efficiency of the propelling mechanism, this scenario would imply a much higher accretion rate at the magnetospheric radius, pushing the magnetospheric edge close to the corotation radius.

Even if no material reaches the WD surface, a reservoir of circumstellar material at the edge of the magnetosphere could produce X-rays, as shocks could develop in the propelling region. Shock-heated plasma or magnetic reconnection at the interaction boundary has been proposed, for example, for the soft X-ray component of the WD propeller AE Aquarii \citep{oruru2012,kitaguchi2014}. Three-dimensional MHD simulations of propellers show that shocks and reconnection layers naturally develop near this boundary, producing luminosities of $L_X\!\sim\!10^{26}$--$10^{28}\,\mathrm{erg\,s^{-1}}$ and power-law indices $\Gamma\!\approx\!2$--3, broadly consistent with our spectral fits for J1901 \citep{wynn1997,romanova_propeller-driven_2005,romanova_mri-driven_2012,romanova_properties_2018,lii_mhd_2014,bozzo_magnetospheric_2018,blinova_comparisons_2019}.

Magnetic reconnection provides another possible source of X-rays, either near the magnetospheric boundary or closer to the WD surface. Turbulent inflows, outflows, or differential rotation could stress and tangle the field lines, leading to repeated reconnection events that heat plasma or accelerate particles. A similar mechanism has been invoked for the pulsed emission of the binary WD pulsar AR~Scorpii \citep{marsh2016,takata2018}. In Ar~Sco, the phase-averaged X-ray spectrum is consistent with optically thin plasma emission, but the isolated pulsed component can be described by a power law with $\Gamma=2.3\pm0.5$ \citep{takata2018}, consistent with the power-law slope measured for J1901. Unlike Ar~Sco, however, J1901 has an X-ray pulse height of $\approx600\%$, so the phase-averaged spectrum is likely dominated by the pulsed component rather than by steady out-of-pulse emission.

Alternatively, if the plasma is supplied by a magnetically driven wind from the white dwarf itself, shocks may form where the wind detaches from the field lines near the magnetospheric edge. Material trapped on closed field lines could also be accelerated towards loop apices, producing shocks analogous to those in magnetically confined winds \citep[see e.g.][]{babel1997,ud-doula2002,ud-doula2008,ud-doula2016}. In this scenario, the X-ray luminosity would depend on the mass-loss rate, the efficiency with which rotational or magnetic energy is transferred to the plasma, and the geometry of the large-scale magnetic field.

The X-ray timing provides important constraints on the emission site. The light curve shows no significant flares or aperiodic variability during the observations, suggesting a relatively steady process. The large pulse height could indicate that much of the emitting region is occulted by the WD, favouring a compact region close to the surface, such as magnetically channelled accretion columns, hot spots, or near-surface reconnection sites. The X-rays are also nearly in anti-phase with the far-UV light curve (Fig.~\ref{fig:lc-comp-normamp}). In our magnetic atmosphere models, the far-UV minimum corresponds to the viewing phase of highest magnetic-field strength (Fig.~\ref{fig:stephane-models-lc}); if accretion or reconnection is concentrated near the magnetic poles, the X-rays would therefore be expected to peak when the highest-field region is in view. This interpretation remains tentative, however, because a self-consistent model of the magnetic geometry and its wavelength-dependent light curves is still required.

If instead the X-rays originate far from the WD surface, occultation by the star cannot explain the strong pulsations. In that case, the modulation would require either a strongly anisotropic emission region or relativistic beaming. Reconnection could accelerate particles to relativistic speeds, and material forced into near-corotation close to the light cylinder could also reach mildly relativistic velocities. Deeper X-ray observations, especially phase-resolved spectroscopy, will be needed to distinguish between a compact surface origin, magnetospheric shocks, and reconnection-powered emission farther from the WD.

\subsection{Comparison with J2008 and other merger remnants}
\label{sec:Comparison}
In our companion paper \citep{cristea2026}, we describe another merger remnant that shows evidence of circumstellar material: ZTF\,J2008+4449 (hereafter J2008). The two systems appear to be close analogues, as their similarities are striking and span several key parameters. Both WDs are very massive (J2008 has a mass of about 1.1\,M$_\odot$) and exhibit short spin periods of approximately 7 minutes (6.9 minutes for J1901, and 6.6 minutes for J2008).
Also, the two objects are among the most highly magnetised WDs known, with field strengths in the range of $5$–$9\times10^8$\,G.

The similar field strengths, rapid rotation, and high masses point to a shared origin as merger remnants. Finally, both J1901 and J2008 are detected as soft X-ray sources, despite being apparently isolated (i.e. they are not in binary systems with mass-transferring companions). Their X-ray spectra are very similar in shape: when fitted by a two-temperature optically thin plasma model, we recover similar temperatures (the soft component is $kT=0.23\pm0.03$\,keV for J1901 and $kT=0.23^{+0.04}_{-0.03}$\,keV for J2008, while the hard component is $kT=2.9^{+1.2}_{-0.8}$\,keV for J1901 and $kT=3.6^{+1.0}_{-0.7}$\,keV for J2008).

The strong similarities hint at a common nature. However, several key differences may indicate distinct evolutionary stages. J2008 is approximately 100 times more X-ray luminous than J1901 (with $L_X \approx 10^{29}$ erg s$^{-1}$ compared with $L_X \approx 1.3 \times 10^{27}$ erg s$^{-1}$ for J1901), and it spins down with a large period derivative of $\dot{P}=(1.80\pm0.09)\times10^{-12}$\,s\,s$^{-1}$ (our current data do not allow us to place a strong constraint on the period derivative of J1901).

Furthermore, J2008 exhibits Balmer (H$\alpha$ and H$\beta$) emission in its optical spectrum, which indicates the existence of a co-rotating half ring of hydrogen-rich material trapped in the magnetosphere close to the WD (at a distance of $\approx$\,20-35 WD radii from the surface). As no emission lines are detected in the spectrum of J1901, if a similar structure is present, the density or temperature in the disc would have to be smaller, causing the emission lines to be too weak to be detected. Finally, J1901 has a cooling age of approximately 470 Myr, while J2008 exhibits a higher effective temperature, implying an almost ten times younger age of about 60 Myr.

\begin{table}[tb]
\def\arraystretch{1.12}
\centering
\caption{Properties of three ultra-massive, highly magnetised WDs.}
\label{tab:wd_comparison}
\resizebox{\columnwidth}{!}{
\begin{tabular}{lccc}
\hline\hline
Property & J2008$^a$ & J1901$^b$ & J0317$^c$ \\
\hline
RA, Dec (deg) & 302.060, 44.826 & 285.386, 14.968 & 49.315, $-85.605$ \\
Distance (pc) & 350 & 41 & 31 \\
$P_{\rm rot}$ (s) & 393 & 416 & 725 \\
$T_{\rm eff}$ (K) & 34,000 & 27,500 & 33,800 \\
$M$ ($M_\odot$) & 1.12 & 1.29 & 1.32 \\
$\log g$ (cgs) & 8.4 & 9.4 & 9.4 \\
$B_{\rm avg}$ (MG) & 500 & 720 & 340 \\
$L_X$ (erg\,s$^{-1}$) & $2.3\times10^{29}$ & $1.4\times10^{27}$ & Non-detection \\
H$\alpha$ emission & Yes & No & No \\
Cooling age (Myr) & $<$100 & 500 & 300\textsuperscript{*} \\
Notes & X-ray, H$\alpha$ & X-ray, pulsed & Rapid rotator \\
\hline
\end{tabular}
}
\tablefoot{J2008 denotes ZTF\,J2008+4449, J1901 denotes ZTF\,J1901+1458, and J0317 denotes RE\,J0317$-$853. The cooling age marked with an asterisk was calculated using the evolutionary models of \citet{althaus_structure_2022} for the parameters reported by \citet{vennes_multiwavelength_2003}. Sources: $^a$\citet{cristea2026}; $^b$this work; $^c$\citet{vennes_multiwavelength_2003,harayama_search_2013,dessert2022}.}
\end{table}

Taken together, these similarities and differences could suggest that J1901 and J2008 represent two snapshots in the life cycle of the same class of systems. J2008 is observed in a younger state, showing stronger X-ray emission and Balmer emission, possibly due to a larger reservoir of circumstellar material.
J1901, older by several hundred million years, has cooled and depleted its circumstellar material, resulting in weaker X-ray emission, the absence of emission lines from trapped material and possibly a smaller spin-down rate. Alternatively, if the X-rays are powered by a wind, the difference in X-ray luminosity could be due to the larger mass and surface gravity of J1901, which would be causing a weaker wind.

For comparison, Table~\ref{tab:wd_comparison} also presents RE J0317-853, another extreme WD combining rapid rotation with high mass and high magnetic-field strength. RE J0317-853 was first identified as an extreme-UV source with ROSAT \citep{shara1993} and characterised by \citet{barstow_re_1995} as a 1.33--1.35\,M$_\odot$ WD possessing a $\approx$340\,MG magnetic field and a 725\,s rotation period, parameters later refined through phase-resolved far-UV spectroscopy, high-precision parallax and spectropolarimetry \citep{ferrario_euve_1997,burleigh_phase-resolved_1999,vennes_multiwavelength_2003}. RE J0317-853, which is a slower rotator with a weaker magnetic field, has a cooling age between those of J1901 and J2008 ($\rm{t}_{\rm cool} \approx200$\,Myr; \citealt{ferrario_euve_1997}). RE J0317-853 was observed for 37\,ks with \textit{Chandra} \citep{safdi2019}, again to search for possible signatures of axion emission. The observation did not yield a detection \citep{dessert2022}. This X-ray non-detection could help distinguish between the potential emission mechanisms discussed above in Section~\ref{sec:x-ray-mechanism}. For instance, when considered alongside the intermediate cooling age, the non-detection makes fallback accretion from merger debris less likely. The comparatively weaker magnetic field and longer spin period also suggest that RE J0317 might not be extreme enough to support outflows from the WD surface. Constraining the X-ray emission from other merger remnants can help pinpoint the origin of the emission in the two newly discovered systems. Several promising candidates fall firmly within the merger-remnant class \citep[see][]{kilic_isolated_2021,kilic_merger_2023,williams84Second1692025}. Although they span somewhat different regions of parameter space in mass, rotation period and magnetic-field strength, none have confirmed X-ray detections, making them compelling targets for dedicated follow-up campaigns.

\begin{table*}
    \centering
    \def\arraystretch{1.25}
	\caption{Summary of measured and derived parameters for ZTF\,J1901+1458.}
	\label{tab:summary}
    \makebox[\textwidth][c]{
    \begin{tabular}{@{} p{0.20\textwidth} p{0.30\textwidth} p{0.30\textwidth} @{}}
            \hline
            \hline
            Origin & Parameter & Value\\
            \hline
            \multirow{3}{*}{From Gaia DR3} &
            Gaia ID & 4506869128279648512 \\
            &Parallax & $24.1538 \pm 0.0489$ mas \\
            & Distance \citet{gaia_collaboration_vizier_2020}  & $41.44 \pm0.08$ pc \\
            \hline
             \multirow{6}{*}{From SED Fitting} &
            Radius of the WD  & $2664^{+113}_{-109}$\,km \\
            &Temperature of the WD & T$_{\rm eff}=27,445^{+680}_{-1390}$\,K \\
            & Mass of the WD & $1.32 \pm 0.01$\,M$_\odot$ [CO], $1.29 \pm 0.01$\,M$_\odot$ [ONe]\\
            & Cooling age of the WD & $0.50 \pm 0.01$\,Gyr [CO], $0.48 \pm 0.01 $\,Gyr [ONe] \\
            & Magnetic Dipole Field Strength & B$_{\rm P} =1,062^{+65}_{-97}$ MG\\
            & Field Inclination & $\theta_{\rm inc}=63^{+11}_{-22}$ deg\\
             [5pt]
            \hline
            \multirow{3}{*}{From light curve analysis} &
            Spin period of the WD & $P_0=6.9373843 \pm 0.0000001$ min \\
            &Reference Ephemeris$^{\rm a}$(MJD) & 58753.38359598\\
            \hline
            \multirow{7}{*}{From X-rays analysis} & X-ray flux [0.25--10.0\,keV] & $7.45_{-0.69}^{+0.68}\times10^{-15}$\,erg\,s$^{-1}$\,cm$^{-2}$\\
            & X-ray luminosity [0.25--10.0\,keV] & {$L_X =(1.44\pm 0.13)\times 10^{27}\,\rm{erg\,s^{-1}}$}  \\
            & Power-law index & $\Gamma=2.43^{+0.17}_{-0.15}$ \\
            &$\dot{\rm M}_{\rm X}$& $\approx$\,2 $\times 10^{9}$\,g\,s$^{-1}$ \\
            & \multirow{2}{*}{2-temperature model} & $kT_1=4.2\pm1.1$\,keV \\
            & & $kT_2=0.22\pm0.03$\,keV \\
            & Pulse fraction & $600\%$ peak/quiescence \\
            \hline
            \hline
	\end{tabular}}
    \tablefoot{$^{\rm a}$ Reference ephemeris used to phase-fold all light curves in this work. The measured X-ray flux and all derived X-ray parameters correspond to the 0.25--10.0\,keV energy band. The 68\% and 90\% confidence limits on these parameters, and those for the 0.25--2.0\,keV soft band, are given in Table~\ref{tab:allinst_counts}. Accretion rates are computed assuming that approximately half of the gravitational potential energy is converted into X-ray luminosity \citep{patterson1985}.}
\end{table*}

\section{Conclusions \label{sec:Conclusions}}

Our multi-wavelength campaign on the ultra-massive, highly magnetised WD ZTF\,J1901+1458 has refined its physical parameters and has characterised the star as an isolated, rapidly spinning merger remnant with high-energy emission powered by the interplay between its rotating magnetosphere and a tenuous supply of circumstellar plasma.

The \textit{HST} UV spectra, analysed with a newly developed set of magnetised atmosphere models, show Zeeman-split Lyman absorption features and a strong spectral break at $\sim$3000\,\AA. Although we currently lack a full grid of models able to properly characterise the structure of the magnetic field of the WD, the new models capture most features of the broadband spectrum. The best-fit parameters, along with estimates of mass and cooling age, are listed in Table~\ref{tab:summary}. Compared with the previous estimate, the star is $\approx$\,25\,$\%$ larger, while placing its central density close to but below the sodium electron-capture threshold. It is therefore unlikely that the WD is currently undergoing Urca cooling in its core, as previously suggested in C21 and \citet{schwab_cooling_2021}. Future phase-resolved modelling with a more complete grid of magnetic atmosphere models will be needed to determine whether the field is predominantly dipolar, contains significant higher-order multi-pole components, or is offset from the centre, as seen in some magnetic WDs \citep[e.g.][]{euchner_zeeman_2002}.

The detection of soft X-rays from an isolated WD reveals a new and unexpected phenomenon. The spectral shape is inconsistent with photospheric emission from the WD, and thus the X-ray emission is likely due to the interaction of the WD or its magnetosphere with circumstellar material. A joint fit to the \textit{Chandra}/ACIS-I and \textit{XMM-Newton}/EPIC spectra favours a single power-law model, although a double-temperature optically thin plasma model is also a good fit to the data (see Table~\ref{tab:summary}).
The 0.25–10\,keV luminosity is $(1.44\pm0.13)\times10^{27}\,$erg\,s$^{-1}$ and the flux is modulated with a $\simeq$\,600\% pulse fraction on the 6.94-minute rotation period. Deeper \textit{Chandra} or \textit{XMM-Newton} observations would enable phase-resolved spectroscopy, allow us to test whether the spectrum changes between pulse and off-pulse phases, improve constraints on the pulse profile, and allow us to  search for long-term changes in the X-ray luminosity.

The near-infrared FIRE spectrum aligns with the WD continuum and shows no excess at 1--2.5\,$\mu$m; any stellar or substellar companion hotter than $\approx$\,700\,K would leave detectable molecular features in the data. By computing irradiation temperatures along brown dwarf evolutionary tracks, we rule out near Roche-lobe-filling brown dwarf companions of all plausible masses and ages, confirming that J1901 is isolated and that the observed X-rays cannot be due to accretion from a binary companion in a cataclysmic variable system. A cold planet or a more distant brown dwarf transferring mass through winds would still be allowed, although unlikely. Deeper infrared observations, particularly with \textit{JWST}/NIRCam or MIRI, could further constrain faint cool substellar companions or thermal emission from material at longer wavelengths.

We have considered three possible sources of circumstellar material, but further studies are needed to verify if any of them are consistent with the observations:
\begin{enumerate}
\item Fallback accretion.
Eccentric merger ejecta may fall back and supply material at late times. Simulations, however, have not yet probed timescales of hundreds of millions of years after the merger, so it remains unclear whether a low level of accretion can persist until the present day.
\item Disrupted planetary body. The tidal disruption of a comet, asteroid, or small planet could be a source of circumstellar material, as is observed in polluted WDs. However, the survival and long-term stability of planetary reservoirs after a double-WD merger remain unexplored.
\item Magnetically driven wind. A weak plasma outflow along open field lines could produce X-rays through shocks or reconnection powered by WD rotation. Winds in this extreme magnetic and rapidly rotating regime have not yet been modelled, and whether a magnetic wind could extract enough material from the surface of the WD to power the observed X-rays remains to be demonstrated.
\end{enumerate}

Because of the rapid rotation and high magnetic field of the WD, any infalling material is likely to be centrifugally expelled once it gets close enough to the star to interact strongly with the magnetic field. In the ejection region, shock-heated plasma could produce the observed X-ray emission. Alternatively, pressure from infalling or outflowing gas on the magnetosphere could trigger magnetic reconnection either near the ejection region or closer to the stellar surface. The large X-ray pulse fraction points to either partial occultation of the emission region by the WD or relativistic beaming, while the clear phase offset between the X-ray and far-UV light curves suggests that at least part of the emission region lies above the surface.

In our companion paper \citep{cristea2026}, we present the discovery of another isolated merger remnant, ZTF\,J2008+4449, with physical properties similar to J1901, including high mass, strong magnetism, rapid rotation, and comparable X-ray emission. The discovery of two rapidly rotating, highly magnetised WDs with high-energy emission opens the possibility of a new class of merger remnants interacting with circumstellar material. Several other promising candidates fall firmly within the merger-remnant class \citep[see][]{kilic_isolated_2021,kilic_merger_2023,williams84Second1692025}. Although these systems span different regions of parameter space in mass, rotation period, and magnetic-field strength, none have yet confirmed X-ray detections, making them compelling targets for dedicated follow-up campaigns. As more merger remnants are identified through time-domain and spectroscopic surveys, future X-ray and infrared observations will be essential for determining how common this phenomenon is, what supplies the circumstellar material, and how magnetic merger remnants evolve.

\begin{acknowledgements}

We thank Lynne Hillenbrand, Srijan Bharati Das, Samarth Hawaldar, CP Johnstone, Alberto Torralba, and Soumyadeep Bhattacharjee for helpful discussions. IC was supported by NASA through grants from the Space Telescope Science Institute, under NASA contracts NASA.22K1813, NAS5-26555 and NAS5-03127. TC was supported by NASA through the NASA Hubble Fellowship grant HST-HF2-51527.001-A awarded by the Space Telescope Science Institute, which is operated by the Association of Universities for Research in Astronomy, Inc., for NASA, under contract NAS5-26555. IT acknowledges the support by Deutsches Zentrum f\"ur Luft- und Raumfahrt (DLR) through grant 50\,OX\,2301. This project has received funding from the European Research Council (ERC) under the European Union’s Horizon 2020 research and innovation programme (Grant agreement No. 101020057). This work was based on observations obtained with the Samuel Oschin Telescope 48-inch and the 60-inch Telescope at the Palomar Observatory as part of the Zwicky Transient Facility project. ZTF is supported by the National Science Foundation under Grants No. AST-1440341, AST-2034437, and currently Award \#2407588. ZTF receives additional funding from the ZTF partnership. Current members include Caltech, USA; Caltech/IPAC, USA; University of Maryland, USA; University of California, Berkeley, USA; University of Wisconsin at Milwaukee, USA; Cornell University, USA; Drexel University, USA; University of North Carolina at Chapel Hill, USA; Institute of Science and Technology, Austria; National Central University, Taiwan, and OKC, University of Stockholm, Sweden. Operations are conducted by Caltech's Optical Observatory (COO), Caltech/IPAC, and the University of Washington at Seattle, USA. This work has made use of data from the European Space Agency (ESA) mission {\it Gaia} (\url{https://www.cosmos.esa.int/gaia}), processed by the {\it Gaia} Data Processing and Analysis Consortium (DPAC, \url{https://www.cosmos.esa.int/web/gaia/dpac/consortium}). Funding for the DPAC has been provided by national institutions, in particular the institutions participating in the {\it Gaia} Multilateral Agreement.

This paper includes data gathered with the 6.5 meter
Magellan Telescopes located at Las Campanas Observatory, Chile.
The Pan-STARRS1 Surveys (PS1) and the PS1 public science archive have been made possible through contributions by the Institute for Astronomy, the University of Hawaii, the Pan-STARRS Project Office, the Max-Planck Society and its participating institutes, the Max Planck Institute for Astronomy, Heidelberg and the Max Planck Institute for Extraterrestrial Physics, Garching, The Johns Hopkins University, Durham University, the University of Edinburgh, the Queen's University Belfast, the Harvard-Smithsonian Center for Astrophysics, the Las Cumbres Observatory Global Telescope Network Incorporated, the National Central University of Taiwan, the Space Telescope Science Institute, the National Aeronautics and Space Administration under Grant No. NNX08AR22G issued through the Planetary Science Division of the NASA Science Mission Directorate, the National Science Foundation Grant No. AST–1238877, the University of Maryland, Eotvos Lorand University (ELTE), the Los Alamos National Laboratory, and the Gordon and Betty Moore Foundation.

This work made use of Astropy,\footnote{http://www.astropy.org} a community-developed core Python package and an ecosystem of tools and resources for astronomy \citep{astropy:2013, astropy:2018, astropy:2022}. This work made use of the perceptually uniform colourmaps from Fabio Crameri \citep{crameri_scientific_2023}. This research was supported by the Scientific Service Units (SSU) of IST Austria through resources provided by Scientific Computing (SciComp).
\end{acknowledgements}

\bibliographystyle{aa}
\bibliography{references}
\clearpage
\begin{appendix}
\onecolumn
\section{Data}
\label{sec:appendix-data}
This Appendix collects supporting data products and methodological details that are used in the main analysis but are not required for the main flow of the paper. The photometry in Table~\ref{tab:photometry} is used in the SED fitting and in the companion constraints discussed in Sections~\ref{sec:Modelling} and \ref{sec:BD}.
\begin{table}[htb]
   \def\arraystretch{1.5}
    \centering
    \caption{Optical and near-infrared photometry of J1901.}
    \label{tab:photometry}
    \begin{tabular}{lccc}
    \hline\hline
    Filter & $\lambda_\mathrm{eff}$ [\AA] & Magnitude & Uncertainty  \\
    \hline
    Pan-STARRS $g$    & 4814  & 15.512 & 0.02 \\
    Pan-STARRS $r$    & 6170  & 15.977 & 0.02 \\
    Pan-STARRS $i$    & 7525  & 16.275 & 0.02 \\
    Pan-STARRS $z$    & 8660  & 16.594 & 0.02 \\
    Pan-STARRS $y$    & 9620  & 16.762 & 0.02 \\
    UKIDSS $J$        & 12350 & 17.339 & 0.11 \\
    UKIDSS $H$        & 16500 & 17.765 & 0.14 \\
    UKIDSS $K_s$        & 21700 & 18.746 & 0.33  \\
    \hline
    \end{tabular}
    \tablefoot{Pan-STARRS and UKIDSS magnitudes are in the AB system.}
\end{table}
Near-infrared images of the WD in the $J$, $H$ and $K_s$ bands are available from the UKIRT Infrared Deep Sky Survey (UKIDSS) \citep{lawrence_ukirt_2007}. The UKIDSS catalogue only provides calibrated images; we therefore performed aperture photometry on the available images to obtain $J$, $H$ and $K_s$ magnitudes for J1901. We first determined the image dimensions and extracted the sky coordinates at the frame edges using the WCS header. A crossmatch with the publicly available Gaia–2MASS catalogue (V/155) \citet{fouesneau_astrophysical_nodate} provided high-accuracy astrometry and proper motion data for stars within the field. The Gaia astrometric solutions were then proper-motion adjusted to the UKIDSS observation epoch, and refined centroids for each source were obtained by fitting 2D Gaussians using \texttt{photutils} \citep[see][]{larry_bradley_2024_13989456}. Aperture photometry was performed at the corrected positions with apertures scaled to the seeing (typically $\sim$\,2$\times$ FWHM), and Poisson noise was propagated through both source and background contributions. To ensure robust flux calibration, we used reference stars with known 2MASS magnitudes and carefully accounted for background subtraction uncertainties using both pixel-based and area-weighted background estimates. The resulting magnitudes and errors are listed in Table~\ref{tab:photometry}.

\section{SED fit and evolutionary tracks}

\begin{figure}[h]
    \centering
    \includegraphics[width=1\linewidth]{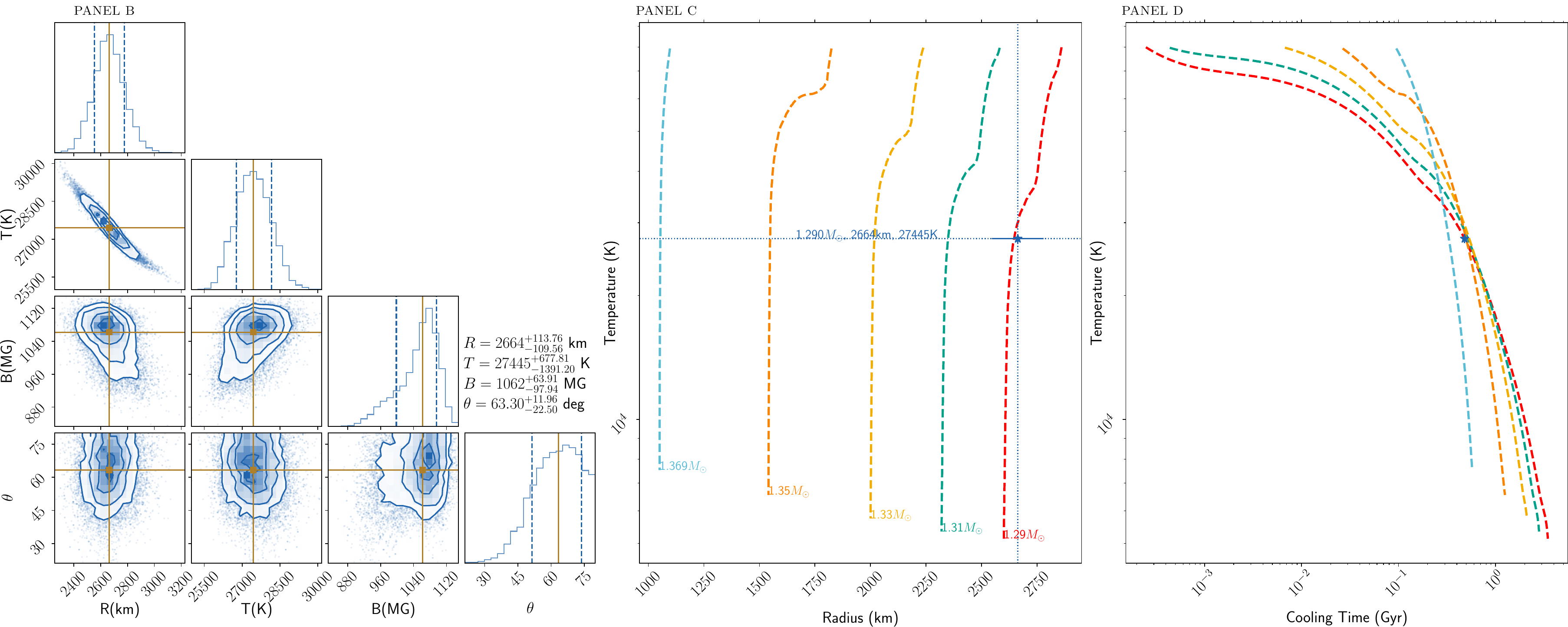}
    \caption{Panel b: Marginalised posterior distributions for the radius, effective temperature, magnetic field strength, and inclination used in the SED fit. The contours indicate the 1 and 3$\sigma$ limits. Panels c,d: ONe-core WD evolutionary tracks for different masses from \citet{althaus_carbon-oxygen_2023}, shown as radius and cooling time vs temperature. The star marks the position of J1901 relative to the evolutionary tracks.}
    \label{fig:cornerplot}
\end{figure}

As shown by the marginalised parameter posteriors in Fig.\,\ref{fig:cornerplot}, \textsc{Panel b}, the radius, temperature, and magnetic field are correlated. Since we fit only the phase-averaged spectrum, this degeneracy is expected. Nonetheless, the magnetic models and the extended \textit{HST} UV coverage significantly improve the characterisation of the WD physical parameters compared to previous estimates.

\section{\textit{HST}, \textit{XMM-Newton}, and \textit{Chandra} observations}
\begin{table}[htb]
\setlength{\tabcolsep}{4pt}
\centering
   \caption{\textit{HST} observations of J1901.}
   \label{tab:HST}
   \begin{tabular}{|c|c|c|c|c|}
   \hline
      Instrument & Mode & Start & End & Duration(s) \\
      \hline
      COS & TIME-TAG & 2022-08-02 14:18:46 & 2022-08-02 14:56:34 & 2268.192 \\
      COS & TIME-TAG & 2022-08-02 15:46:05 & 2022-08-02 16:30:58 & 2693.184 \\
      COS & TIME-TAG & 2022-09-17 10:22:44 & 2022-09-17 11:00:32 & 2268.192 \\
      COS & TIME-TAG & 2022-09-17 11:50:02 & 2022-09-17 12:34:55 & 2693.216 \\
      \hline
      \hline
      STIS & TIME-TAG & 2023-08-19 02:21:44 & 2023-08-19 02:57:07 & 2122.952 \\
      STIS & TIME-TAG & 2023-08-19 03:46:22 & 2023-08-19 03:52:37 & 374.639 \\
      STIS & TIME-TAG & 2023-08-19 05:22:07 & 2023-08-19 06:06:31 & 2664.097 \\
      STIS & TIME-TAG & 2023-09-10 09:34:32 & 2023-09-10 10:18:56 & 2664.097 \\
      STIS & TIME-TAG & 2023-09-10 11:09:35 & 2023-09-10 11:53:59 & 2664.097 \\
      STIS & TIME-TAG & 2024-06-23 10:08:55 & 2024-06-23 10:42:13 & 1998.072 \\
   \hline
   \end{tabular}
   \tablefoot{The total COS and STIS exposure times were 9922.784\,s and 12023.952\,s, respectively.}
\end{table}

\begin{table}[htb]
\def\arraystretch{1.05}
   \centering
   \caption{\textit{XMM-Newton} and \textit{Chandra} observations of J1901.}
   \label{tab:X-ray}
   \begin{tabular}{|c|c|c|c|c|c|}
   \hline
      Instrument & Mode & Filter & Start & End & Duration(s) \\
      \hline
      \multicolumn{6}{|c|}{\textit{XMM} EPIC} \\
      \hline
      MOS1 & Large Window & THIN1 & 2024-03-21 13:51:28 & 2024-03-22 10:02:02 & 72634 \\
      MOS2 & Large Window & THIN1 & 2024-03-21 13:51:53 & 2024-03-22 10:05:17 & 72804 \\
      pn   & Full Frame   & THIN1 & 2024-03-21 14:17:10 & 2024-03-22 10:26:30 & 72560 \\
      \hline
      \multicolumn{6}{|c|}{XMM OM} \\
      \hline
      OM & Fast & UVW1 & 2024-03-21 13:59:50 & 2024-03-21 15:13:10 & 4400 \\
      OM & Fast & UVW1 & 2024-03-21 15:33:17 & 2024-03-21 16:46:37 & 4400 \\
      OM & Fast & UVW1 & 2024-03-21 16:51:44 & 2024-03-21 18:05:04 & 4400 \\
      OM & Fast & UVW1 & 2024-03-21 18:10:11 & 2024-03-21 19:23:32 & 4401 \\
      OM & Fast & UVW1 & 2024-03-21 19:28:38 & 2024-03-21 20:41:57 & 4399 \\
      OM & Fast & UVW1 & 2024-03-21 20:47:05 & 2024-03-21 22:00:25 & 4400 \\
      OM & Fast & UVW1 & 2024-03-21 22:05:32 & 2024-03-21 23:18:51 & 4399 \\
      OM & Fast & UVW1 & 2024-03-21 23:24:00 & 2024-03-22 00:05:39 & 2499 \\
      \hline
      OM & Fast & UVM2 & 2024-03-22 00:10:46 & 2024-03-22 01:24:05 & 4399 \\
      OM & Fast & UVM2 & 2024-03-22 01:29:13 & 2024-03-22 02:42:33 & 4400 \\
      OM & Fast & UVM2 & 2024-03-22 02:47:41 & 2024-03-22 04:00:59 & 4398 \\
      OM & Fast & UVM2 & 2024-03-22 04:06:07 & 2024-03-22 05:19:27 & 4400 \\
      OM & Fast & UVM2 & 2024-03-22 05:24:34 & 2024-03-22 06:37:54 & 4400 \\
      OM & Fast & UVM2 & 2024-03-22 06:43:02 & 2024-03-22 07:56:20 & 4398 \\
      OM & Fast & UVM2 & 2024-03-22 08:01:27 & 2024-03-22 09:14:46 & 4399 \\
      OM & Fast & UVM2 & 2024-03-22 09:19:54 & 2024-03-22 10:33:14 & 4400 \\
      \hline
      \multicolumn{6}{|c|}{\textit{Chandra}} \\
      \hline
      ACIS-I & VFAINT/TE & - & 2022-12-09 14:35:08 & 2022-12-09 16:56:39 & 14910 \\
      ACIS-I & VFAINT/TE & - & 2022-12-10 00:38:11 & 2022-12-10 02:58:31 & 14420 \\
      ACIS-I & VFAINT/TE & - & 2022-12-10 10:33:10 & 2022-12-10 12:11:29 & 9980 \\
      \hline
   \end{tabular}
   \tablefoot{The \textit{XMM-Newton} observations comprise EPIC and OM data (ObsID 0922750101; PI Traulsen), while the \textit{Chandra} observations used ACIS-I (proposal ID 24200244). The total EPIC exposures were 72.6\,ks for MOS1, 72.8\,ks for MOS2, and 72.5\,ks for pn. The total OM exposures were 32.9\,ks in UVW1 and 35.3\,ks in UVM2; the total ACIS-I exposure was 39.3\,ks.}
\end{table}
\clearpage
\section{X-ray counts}
\begin{table}[ht]
\def\arraystretch{1.05}
\centering
\caption{\textit{XMM-Newton} EPIC event counts and detection statistics.}
\begin{tabular}{l l r r r r r}
\hline
\hline
Instr. & Band & src & bkg & CL$_{90}^{\rm low}$ & CL$_{90}^{\rm high}$ & Sig. \\
Instr. & [keV] & [cts] & [exp. cts] & \multicolumn{2}{c}{[$10^{-4}$ cts\,s$^{-1}$]} & \multicolumn{1}{c}{$\sigma$} \\
\hline
\multirow{4}{*}{pn} & 0.2--0.5 & 117 & 46.1 & 10.2 & 16.9 & 10.45 \\
 & 0.5--2.0 & 147 & 69.3 & 11.0 & 18.6 & 9.33 \\
 & 2.0--4.5 & 42 & 24.7 & 1.4 & 5.4 & 3.49 \\
 & 4.5--7.5 & 31 & 24.2 & 0.0 & 2.9 & 1.37 \\
 & 0.2--7.5 & 306 & 140.1 & 26.0 & 36.8 & 14.02 \\
\hline
\hline
\multirow{4}{*}{M1} & 0.2--0.5 & 17 & 9.7 & 0.2 & 2.3 & 2.33 \\
 & 0.5--2.0 & 48 & 21.5 & 2.5 & 6.0 & 5.72 \\
 & 2.0--4.5 & 13 & 10.0 & 0.0 & 1.4 & 0.96 \\
 & 4.5--7.5 & 5 & 6.8 & 0.0 & 0.7 & -- \\
 & 0.2--7.5 & 78 & 41.2 & 3.6 & 8.0 & 5.74 \\
\hline
\hline
\multirow{4}{*}{M2} & 0.2--0.5 & 31 & 7.3 & 2.3 & 5.0 & 8.75 \\
 & 0.5--2.0 & 60 & 23.9 & 3.6 & 7.4 & 7.39 \\
 & 2.0--4.5 & 14 & 8.6 & 0.1 & 1.8 & 1.84 \\
 & 4.5--7.5 & 8 & 9.0 & 0.0 & 0.8 & -- \\
 & 0.2--7.5 & 105 & 39.8 & 7.3 & 12.3 & 10.33 \\
\hline
\hline
\end{tabular}
\tablefoot{The net source (src) counts were measured in the 20 arcsec source aperture without background subtraction. The expected background (bkg) counts are the counts measured in the 80 arcsec background aperture and scaled to the source-aperture area. The 90\% confidence limits (CL) on the source count rates are shown with the associated detection significance. The pn, M1, and M2 effective exposure times after selecting good time intervals were 53.1, 65.1, and 67.3\,ks, respectively.}
\label{tab:allinst_counts}
\end{table}

\begin{table}[ht]
\def\arraystretch{1.05}
\centering
\caption{\textit{Chandra} ACIS-I event counts and detection statistics.}
\begin{tabular}{l l r r r r r}
\hline
\hline
Instr. & Band & src & bkg & CL$_{90}^{\rm low}$ & CL$_{90}^{\rm high}$ & Sig. \\
Instr. & [keV] & [cts] & [exp. cts] & \multicolumn{2}{c}{[$10^{-5}$ cts\,s$^{-1}$]} & \multicolumn{1}{c}{$\sigma$} \\
\hline
\multirow{6}{*}{ACIS-I} & 0.2--0.5 & 0 & 0.01 & 0.0 & 5.9 & -- \\
 & 0.5--1.2 & 1 & 0.05 & 0.1 & 9.8 & 1.6 \\
 & 1.2--2.0 & 6 & 0.06 & 6.9 & 28.0 & 6.5 \\
 & 0.5--2.0 & 7 & 0.11 & 8.5 & 31.2 & 6.5 \\
 & 2.0--7.0 & 3 & 0.23 & 1.8 & 17.1 & 2.9 \\
 & 0.5--7.0 & 10 & 0.33 & 13.5 & 40.4 & 6.9 \\
\hline
\hline
\end{tabular}
\tablefoot{The net source (src) counts were measured in the 2.0 arcsec source aperture without background subtraction. The expected background (bkg) counts are the counts measured in the 80 arcsec background aperture and scaled to the source-aperture area. The 90\% confidence limits (CL) on the source count rates are shown with the associated detection significance. The total ACIS-I exposure time was 39.3\,ks.}
\label{tab:chandra_acis_counts}
\end{table}

\begin{figure*}[htb]
    \centering
    \includegraphics[width=0.38\linewidth]{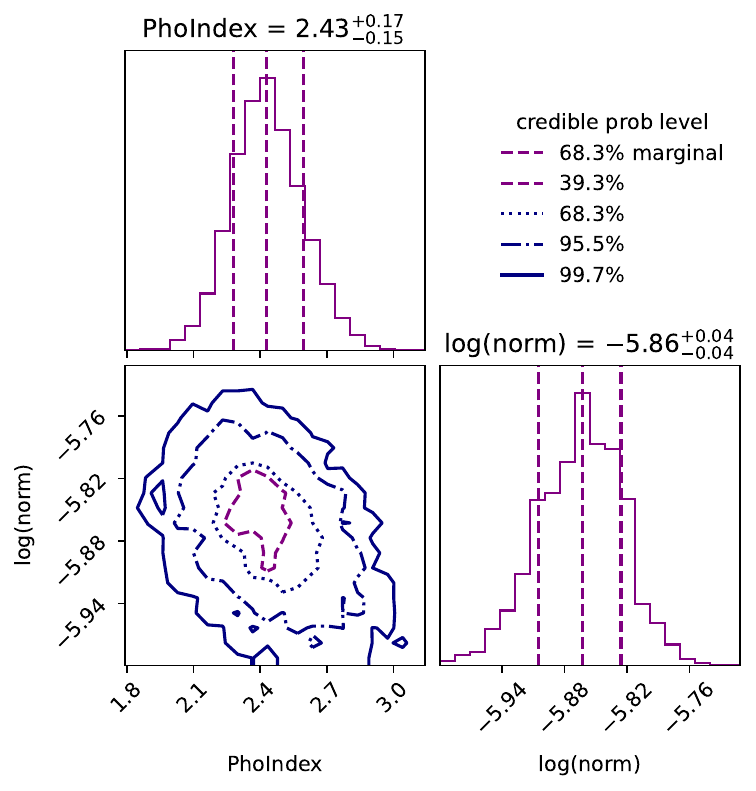}
    \includegraphics[width=0.61\linewidth]{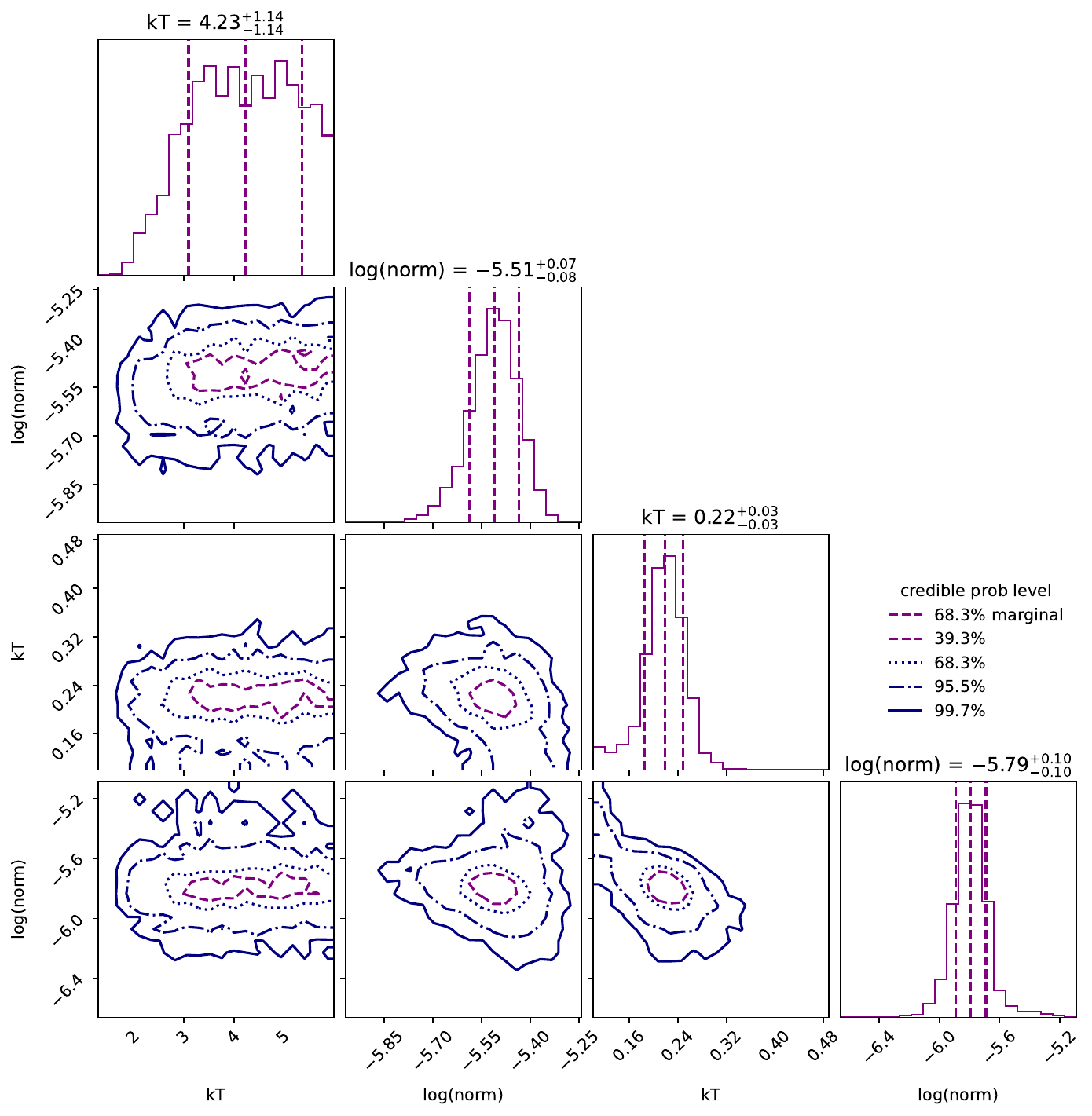}
    \caption{Left: Posterior distributions of the two parameters included in the X-ray spectral fitting of the power-law model. Right: Posterior distributions of the four parameters included in the X-ray spectral fitting of the two-temperature, optically thin plasma model.}
    \label{fig:X-ray-corner-apec-apec}
\end{figure*}
\vspace{20em}
\end{appendix}
\end{document}